# Risk and Protective Factors in Parkinson's Disease

Iman Beheshti

Department of Human Anatomy and Cell Science, Rady Faculty of Health Sciences, University of Manitoba, Winnipeg, MB, Canada

Corresponding Author:
Dr. Iman Beheshti
Address: 745 Bannatyne Ave., Winnipeg, Manitoba, R3E 0J9, Canada
Email: Iman.Beheshti@umanitoba.ca

https://scholar.google.ca/citations?user=x0AqYvkAAAAJ&hl=en&oi=ao

**Abstract:**

Understanding the risk and protective factors associated with Parkinson's disease (PD) is crucial for improving outcomes for patients, individuals at risk, healthcare providers, and healthcare systems. Studying these factors not only enhances our knowledge of the disease but also aids in developing effective prevention, management, and treatment strategies. This paper reviews the key risk and protective factors associated with PD, with a particular focus on the biological mechanisms underlying these factors. Risk factors include genetic mutations, racial predispositions, and environmental exposures, all of which contribute to an increased likelihood of developing PD or accelerating disease progression. Conversely, protective factors such as regular physical exercise, adherence to a Mediterranean diet, and higher urate levels have demonstrated potential to reduce inflammation and support mitochondrial function, thereby mitigating disease risk. However, identifying and validating these factors presents significant challenges. These include the absence of reliable biomarkers, the intricate interactions between genetic and environmental components, and the clinical heterogeneity observed in PD patients. These barriers complicate the establishment of clear causal relationships and hinder the development of targeted prevention strategies. To overcome these challenges, we propose several solutions and recommendations. Future research should prioritize the development of standardized biomarkers for early diagnosis, investigate gene-environment interactions in greater depth, and refine animal models to better mimic human PD pathology. Additionally, we offer actionable recommendations for PD prevention and management, tailored to healthy individuals, patients diagnosed with PD, and healthcare systems. These strategies aim to improve clinical outcomes, enhance quality of life, and optimize healthcare delivery for PD.

## 1- Introduction

Parkinson's disease (PD) is a progressive neurodegenerative disorder characterized by motor disabilities, making it the second leading cause of such disabilities in adults after strokes. The disease is marked by classical motor manifestations such as bradykinesia, rigidity, and gait disorders, which fluctuate as the disease progresses [1]. At its core, PD involves the degeneration of dopaminergic neurons in the substantia nigra pars compacta, leading to a decrease in dopamine levels in the striatum and the formation of intracellular aggregates known as Lewy bodies [1]. The disease's etiology involves both genetic and environmental factors, which are crucial to understanding and developing effective treatments and prevention strategies [1-5]. Genetic research has identified risk factors such as mutations in genes like LRRK2 and GBA1, which are common in PD patients, suggesting a genetic predisposition to the disease. These mutations vary across populations, offering potential for more tailored treatments [3]. Environmental factors, including pesticides, heavy metals, and air pollution, may contribute to oxidative stress, mitochondrial dysfunction, and neuroinflammation, which are key mechanisms in PD progression [6]. Lifestyle factors can also play a protective role. Regular exercise, especially high-intensity aerobic activity, has been shown to preserve dopamine-producing neurons and slow symptom progression [7]. Diets rich in antioxidants, polyphenols, and polyunsaturated fatty acids may reduce PD risk, and cognitive and social engagement has been linked to protection against cognitive decline in PD patients, emphasizing the importance of mental stimulation [8]. Recent studies have also focused on the gut-brain axis and microbiome dysbiosis, offering new therapeutic possibilities [9, 10].

In this review, we explore various categories of risk and protective factors associated with PD, focusing primarily on the biological mechanisms behind these interactions, rather than just the epidemiological correlations. We address the current challenges in identifying and verifying these factors, and propose potential solutions, such as the development of standardized biomarkers, studying gene-environment interactions, and improving animal models. Importantly, we also emphasize the need for increased awareness and education for patients, at-risk individuals, and healthcare providers—an aspect often overlooked in other research. This includes providing insight into the underlying molecular mechanisms of risk and protective factors in PD. By integrating genetic, environmental, and lifestyle factors, our research aims to enhance the understanding of PD's risk and protective mechanisms. Finally, we highlight preventive strategies, such as lifestyle changes and early interventions, to reduce the global burden of PD and improve patient outcomes.

## 2- Molecular Mechanisms of PD

The molecular mechanisms underlying PD constitute a complex and interconnected network. Key factors, including α-synuclein aggregation, mitochondrial dysfunction, oxidative stress, neuroinflammation, ferroptosis, and gut dysbiosis, all contribute to the progressive neurodegeneration seen in PD (Fig 1). Understanding these mechanisms is essential for identifying risk factors, preventive strategies, and potential interventions to slow or halt PD progression. Table 1 presents a list of the key biological mechanisms in PD along with their associated risk factors. This section provides an overview of the primary molecular processes involved in PD, highlighting their interactions and roles in the pathogenesis of the disease.

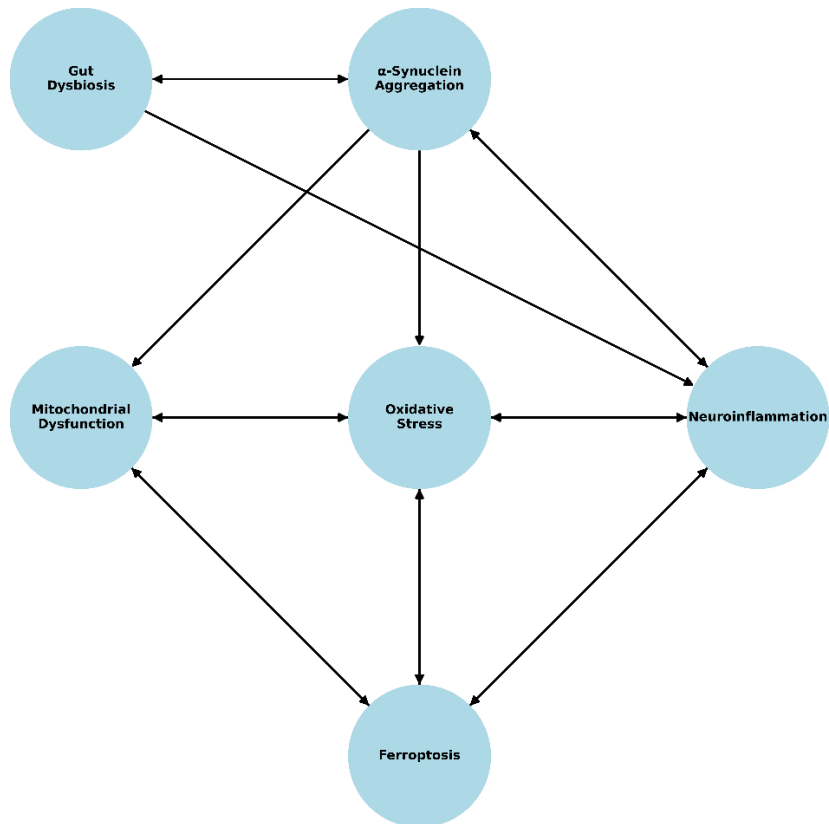

**Figure 1:** Interrelationship of Key Biological Mechanisms in PD.

### 2.1 α-Synuclein Aggregation

α-Synuclein aggregation is considered one of the earliest and most central events in PD pathogenesis. Under pathological conditions, α-synuclein, a presynaptic neuronal protein, misfolds and aggregates into insoluble fibrils, forming Lewy bodies—a hallmark of PD [11]. These aggregates disrupt neuronal function and are believed to propagate from cell to cell, contributing to the spread of pathology. Recent studies have shown that α-synuclein aggregates can be transmitted to neighboring cells, seeding further aggregation in healthy neurons. This cell-to-cell transmission is a critical factor in the spread of pathology in PD [12]. Furthermore, the binding of calcium ions (Ca2+) to the C terminus of α-synuclein has been shown to regulate its binding to synaptic membranes, which may influence aggregation dynamics [13]. α-Synuclein aggregation initiates a cascade of events that lead to:

- Mitochondrial dysfunction through direct interaction with mitochondria, impairing their function and reducing ATP production.
- Oxidative stress by disrupting cellular metabolism and mitochondrial function.
- Neuroinflammation by activating microglia, the brain's immune cells.
- Gut dysbiosis, as misfolded α-synuclein, can potentially propagate through the gut-brain axis.

**2.2 Mitochondrial Dysfunction**

Mitochondrial dysfunction plays a pivotal role in PD pathogenesis. Impairment of mitochondrial function, particularly of complex I of the electron transport chain, leads to reduced ATP production and increased generation of reactive oxygen species (ROS)[14]. This dysfunction is exacerbated by the interaction of α-synuclein aggregates with mitochondria. Recent research has highlighted a generalized reduction in mitochondrial quality control proteins in dopaminergic neurons from PD patients. This includes a decrease in proteins such as PINK1, Parkin, and mitochondrial chaperones, which are crucial for maintaining mitochondrial proteostasis and function [14]. Mitochondrial dysfunction contributes to:

- Oxidative stress through increased ROS production.
- Energy deficits in neurons, particularly affecting dopaminergic neurons in the substantia nigra.

- Ferroptosis by enhancing iron accumulation and lipid peroxidation.

## 2.3. Oxidative Stress

Oxidative stress occurs when there's an imbalance between the production of ROS and the body's ability to detoxify these reactive intermediates or repair the resulting damage. In PD, elevated ROS levels cause damage to lipids, proteins, and DNA, exacerbating neuronal injury. The interplay between oxidative stress and other PD mechanisms creates a vicious cycle:

- Mitochondrial dysfunction is both a cause and consequence of oxidative stress.
- Oxidative damage activates microglia, inducing neuroinflammation.
- Lipid peroxidation caused by oxidative stress contributes to ferroptosis of dopaminergic neurons.

## 2.4. Neuroinflammation

Neuroinflammation in PD involves the activation of microglia and other immune cells in response to neuronal damage and α-synuclein aggregates. This chronic inflammation further damages neurons and promotes disease progression. Recent studies have shown that microglial activation is not only a response to neuronal damage but also a contributor to the progression of PD [15]. Activated microglia can release exosomes containing α-synuclein, which may propagate neuroinflammation and neurodegeneration by affecting distant dopaminergic neurons. Neuroinflammation exacerbates PD pathology by:

- Promoting further α-synuclein aggregation.
- Increasing oxidative stress through the release of pro-inflammatory cytokines and ROS.
- Contributing to ferroptosis by increasing the iron burden in neurons.

## 2.5. Ferroptosis

Ferroptosis is an iron-dependent form of regulated cell death driven by lipid peroxidation. In PD, iron accumulation in the substantia nigra enhances oxidative stress and lipid peroxidation, leading to ferroptotic death of dopaminergic neurons. Recent research has linked α-synuclein aggregation to ferroptosis [16]. Misfolded α-synuclein can disrupt cellular iron homeostasis and promote lipid peroxidation, thereby facilitating ferroptosis. Ferroptosis contributes to PD progression by:

- Exacerbating oxidative stress through excessive ROS production and lipid peroxidation.
- Triggering neuroinflammation through the release of damage-associated molecular patterns (DAMPs).
- Further impairing mitochondrial function due to iron accumulation and oxidative stress.

## 2.6. Gut Dysbiosis

Emerging evidence suggests that gut dysbiosis, an imbalance in the gut microbiota composition, may precede central nervous system involvement in PD. The gut-brain axis allows for the potential propagation of misfolded α-synuclein from the enteric nervous system to the brain via the vagus nerve. Studies have shown that individuals with PD exhibit significant alterations in gut microbiota composition compared to healthy controls. There is a notable depletion of short-chain fatty acid (SCFA)-producing bacteria and an enrichment of pro-inflammatory bacteria [17]. Gut dysbiosis contributes to PD pathology by:

- Facilitating the propagation of α-synuclein aggregates from the gut to the brain.
- Activating systemic immune responses and promoting neuroinflammation.
- Altering the production of neurotransmitters and metabolites that influence brain function.

**Table 1:** Main Biological Mechanisms and Associated Factors in PD.

| Mechanism | Factors |
|---|---|
| α-Synuclein Aggregation | - **Genetics:** SNCA gene mutation (early-onset PD, Lewy body formation) |
| | - **Environmental:** Heavy metals (mercury, manganese, lead) |
| | - **Infections:** Viral infections (trigger immune responses) |
| | - **Gut Dysbiosis:** Disrupts gut-brain axis, promotes misfolding |
| | - **Non-Motor Symptoms:** Anosmia, REM sleep behavior disorder |
| Oxidative Stress | - **Aging:** Natural increase in oxidative damage |
| | - **Genetics:** PARK7 (DJ-1) mutation (oxidative stress defense failure) |
| | - **Environmental:** Pesticides, herbicides, heavy metals, air pollution |
| | - **Metabolic Factors:** Low LDL-C (loss of antioxidant protection) |
| | - **Mixed Effects:** Hyperuricemia (antioxidant but linked to decline) |
| Ferroptosis | - **Aging:** Iron accumulation in substantia nigra |
| | - **Heavy Metals:** Iron, manganese, lead (redox imbalance) |
| | - **Dietary Factors:** Dairy consumption (potential increased iron load) |
| | - **Industrial Chemicals:** Solvents (trichloroethylene) |
| Mitochondrial Dysfunction | - **Genetics:** PINK1, PRKN (Parkin) mutations (defective mitophagy) |
| | - **Environmental:** Herbicides, pesticides, heavy metals, solvents |
| | - **Metabolic Factors:** Insulin dysregulation, diabetes medications |
| Neuroinflammation | - **Aging:** Increased pro-inflammatory cytokines |
| | - **Sex Differences:** Men at higher risk (estrogen is neuroprotective) |
| | - **Genetics:** PARK7 (DJ-1) gene (linked to inflammation) |
| | - **Infections:** Viral infections, chronic inflammatory diseases |
|  | - **Environmental:** Heavy metals, air pollution |
|  | - **Psychological:** Chronic stress, depression, anxiety (HPA axis) |
| **Gut Dysbiosis** | - **Gut-Brain Axis:** Links gastrointestinal and neurodegenerative issues |
|  | - **Chronic Constipation:** Early symptom, linked to PD progression |
|  | - **Antibiotic Use:** Macrolides, lincosamides (gut microbiome disruption) |
|  | - **Probiotics/Diet:** Possible protective role in PD |

## 3- Risk Factors in PD

Risk factors are conditions that increase the likelihood of developing a disease. These can be genetic, environmental, or lifestyle-related and may be modifiable (e.g., smoking, diet) or non-modifiable (e.g., age, family history). For PD, risk factors include age (more common in those over 60), genetics (family history), environmental toxins (e.g., pesticides), and head injuries (Figure 2). Disease progression is slower in those diagnosed at a younger age, while severe motor symptoms, cognitive decline, and poor response to treatments suggest faster progression. Non-motor symptoms like depression and sleep issues also affect disease course.

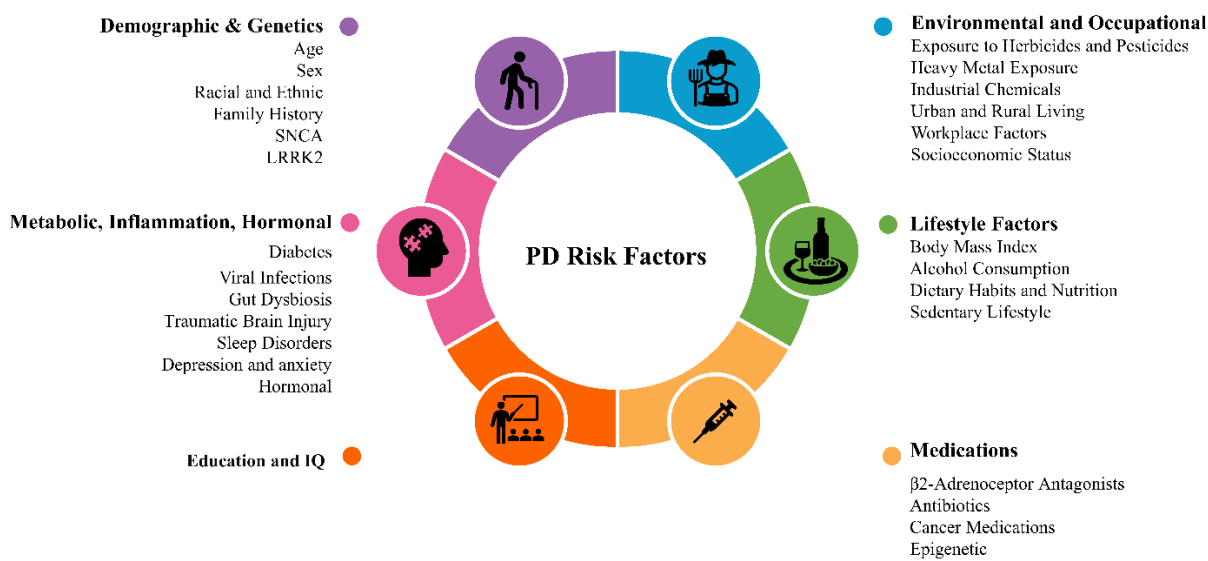

**Figure 2:** Classification of Various Risk Factors in Parkinson's Disease

### 3.1 Demographic and Genetic Factors

Demographic and genetic factors are key risk factors in PD, as they significantly impact the disease's onset, progression, and response to treatment. An understanding of these factors is essential for enhancing diagnosis, improving treatment outcomes, and predicting disease trajectories, thereby enabling the customization of interventions to meet individual patient needs (Table 2).

**Age**

Age is the most significant risk factor for PD, with a median onset age of 60 years [1, 5]. Several mechanisms explain this association. Aging is linked to neuronal loss, with approximately 30% of elderly non-PD individuals experiencing degeneration in the substantia nigra (SN) [18]. This neuronal loss heightens the sensitivity of SN neurons to mitochondrial dysfunction and protein degradation impairments [5]. PD symptoms manifest when 50-60% of SN neurons are lost. Aging also contributes to increased oxidative stress, which damages DNA, proteins, and lipids, accelerating dopaminergic neuron death. Iron accumulation with age generates free radicals via the Fenton reaction, exacerbating oxidative stress and neurodegeneration [19]. Furthermore, aging reduces the efficiency of autophagy and the ubiquitin-proteasome system, processes critical for protein quality control, thereby facilitating the accumulation of misfolded proteins like α-synuclein, a hallmark of PD [20]. Mitochondrial dysfunction due to aging further diminishes energy production and increases oxidative damage. While PD predominantly affects older individuals, about 5% of cases occur in those under 50 years, often due to genetic mutations [21] [5].

**Sex**

PD is approximately twice as prevalent in men compared to women [21]. Protective effects of female sex hormones, such as estrogen, may explain this disparity. Estrogen's neuroprotective roles include upregulating neurotrophic factors, enhancing dopamine synthesis, reducing inflammation, and preventing α-synuclein aggregation and Lewy body formation [22]. Reduced lifetime estrogen exposure, as seen with early menopause or oophorectomy, correlates with higher PD risk [22, 23]. Sex-specific genetic mechanisms also influence PD risk. Variants like the LRRK2 G2019S mutation exhibit higher prevalence in women and sex-dependent phenotypic differences [22]. Urate levels, an endogenous antioxidant, are associated with reduced PD risk and slower progression, particularly in men [23]. Differences in immune response and neuroinflammation may further contribute to the higher PD incidence in men [22].

**Racial and Ethnic Groups**

The prevalence of PD varies across racial and ethnic groups, with Caucasians and Hispanics having higher risks compared to Black and Asian populations [21, 24]. Black individuals show approximately half the PD prevalence of White individuals [25]. This disparity is attributed to genetic, biological, and socioeconomic factors. Genetic predispositions vary among ethnicities, with certain mutations being population-specific. For instance, the LRRK2 G2019S variant is prevalent in Europe, North Africa, and the United States but rare in Japan, where LRRK2 p.R1628P and p.G2385R variants are more common [26]. Ancestry-related gene expression differences also influence PD risk. Ancestry-associated differentially expressed genes (DEGs) account for up to 27% of PD heritability and often relate to immune responses and vascular functions [27]. Environmental and socioeconomic factors intersect with genetic predispositions, influencing PD progression. Stress, healthcare access, and comorbidities play critical roles in exacerbating disparities [28]. Racial differences in immune response and neuroinflammation—evident in ancestry-associated gene expression patterns—may further explain variability in PD risk and progression [27].

**Family History of PD**

A family history of PD is a significant risk factor, with 15-25% of individuals with PD reporting familial connections, and 10-20% having a first-degree relative with the condition [29, 30]. Individuals with a family history are up to three times more likely to develop PD, often presenting with an earlier onset of symptoms [2, 30]. Interestingly, a familial background of melanoma is also linked to an increased risk of PD [31]. Familial PD accounts for approximately 15% of cases, with the remaining 85% categorized as sporadic [32].

**Genetic Contributions**

Approximately 5% of PD cases result from mutations in specific genes, with over 200 PD-related genes identified to date [33]. Key genetic contributors include:

**SNCA**: Mutations in the SNCA gene cause alpha-synuclein aggregation into Lewy bodies, a hallmark of PD. SNCA mutations are often linked to early-onset PD (mean onset: 46 years) and cognitive decline [3, 34].

**LRRK2**: Mutations in the LRRK2 gene are the most common cause of familial PD and also increase sporadic PD risk. LRRK2 mutations, such as G2019S, disrupt autophagy and synaptic function, leading to dopamine toxicity [35-39].

**GBA1**: Mutations in the GBA1 gene, which encodes glucocerebrosidase, impair lysosomal function and are associated with both familial and sporadic PD. They account for 7-10% of PD cases [39-41].

**PARK7 (DJ-1)**: Mutations in PARK7 disrupt oxidative stress defense and neuroinflammatory responses, contributing to early-onset PD (<50 years) and potentially idiopathic PD [42-44].

**PINK1**: Mutations in PINK1 impair mitochondrial function and autophagy, leading to hereditary early-onset PD [45-49].

**PRKN (Parkin)**: Mutations in PRKN affect lysosomal activity, leading to toxic protein aggregation and dopaminergic neuron loss. PRKN mutations are linked to 10-20% of early-onset PD cases [50-54].

**UCHL1**: Abnormalities in UCHL1 disrupt the ubiquitin-proteasome system, contributing to protein aggregation and motor symptom severity in PD [55-58].

**Double Mutations and Combined Genetic Risks**:

In some cases, individuals may carry multiple genetic mutations, such as LRRK2 G2019S combined with GBA1 mutations, which can lead to earlier onset and faster progression of PD symptoms. These combinations can also influence non-motor symptoms like dementia, psychosis, and REM sleep behavior disorder (RBD) [59-62].

**Table 2:** Overview of Demographic and Genetic Factors in PD.

| Category | Subcategory | Key Observations | Mechanisms | References |
|---|---|---|---|---|
| **Age** | | Median onset at 60 years; ~5% cases under 50 | Neuronal loss in substantia nigra; increased oxidative stress; mitochondrial dysfunction; iron accumulation | [5, 18, 19] |
| **Sex** | | Twice as prevalent in men compared to women | Protective effects of estrogen; sex-specific genetic mechanisms (e.g., LRRK2 variants); differences in immune response | [22, 23] |
| **Racial/ Ethnic** | | Higher prevalence in Caucasians and Hispanics; lower in Black and Asian populations | Genetic predispositions; ancestry-linked gene expression; environmental and socioeconomic interactions | [25, 27] |
| **Gene** | SNCA | Associated with early-onset PD and cognitive decline | Alpha-synuclein aggregation; Lewy body formation | [3, 34]. |
| | LRRK2 | Most common cause of familial PD; also linked to sporadic PD | Disrupted autophagy; synaptic dysfunction; dopamine toxicity | [35-39] |
| | GBA1 | Found in 7-10% of PD cases; linked to familial and sporadic forms | Impaired lysosomal function; glucosylceramide buildup | [39-41] |
| | PARK7 (DJ-1) | Associated with early-onset and idiopathic PD | Oxidative stress defense disruption; neuroinflammation | [42-44] |
| | PINK1 | Linked to hereditary early-onset PD | Mitochondrial dysfunction; impaired autophagy | [45-49] |
| | PRKN (Parkin) | Present in 10-20% of early-onset PD cases | Lysosomal dysfunction; toxic protein aggregation | [50-54] |
| | UCHL1 | Associated with protein aggregation and motor symptom severity | Disruption of ubiquitin-proteasome system; alpha-synuclein buildup | [55-58] |

GBA: Glucosylceramidase Beta; LRRK2: Leucine-Rich Repeat Kinase 2; PARK7: Parkin RBR E3 Ubiquitin Protein Ligase; PINK1: PTEN-induced kinase 1; PRKN: Parkin; SNCA: Alpha-synuclein; UCHL1: Ubiquitin C-Terminal Hydrolase L1.

## 3.2 Education and Intelligence

Surprisingly, a high level of education (9 or more years), as well as a higher cognitive capacity in early adulthood, measured by intelligence (IQ), are associated with a higher risk of developing PD, particularly among men [63-65].While the reasons for this link remain unclear, several theories have been proposed. Firstly, lifestyle differences may play a role. Research indicates that individuals with higher IQs or more education often have lower cholesterol levels [66], which have been associated with a higher risk of PD [67, 68]. This relationship between IQ, education, and cholesterol levels might also be influenced by socioeconomic status and lifestyle habits [66]. Additionally, individuals with higher IQs or education levels are less likely to smoke [69, 70], and as previously noted, smoking is considered protective against PD. Occupational differences may also contribute; people with higher education levels often engage in jobs requiring less physical activity, potentially increasing PD risk [71]. Conversely, adults with lower education levels are more likely to smoke, possibly due to stress or targeted tobacco advertising [65]. However, higher education is associated with greater cognitive reserve, which may influence cognitive performance in PD and potentially delay its progression [72]. Table 3 presents a summary of the relationship between education, IQ, and the risk of developing PD.

**Table 3:** Summary of Relationship Between Education, IQ, and PD.

| Category | Key Observations | Mechanisms | References |
|---|---|---|---|
| **Education & IQ** | Higher education (≥9 years) and IQ associated with increased PD risk, particularly in men | Lower cholesterol levels; reduced smoking rates; less physical activity; higher cognitive reserve | [70-72] |
| **Lifestyle Factors** | Individuals with higher education/IQ exhibit healthier lifestyles (e.g., non-smoking, lower cholesterol) | Reduced smoking may negate protective effects; lower cholesterol linked to increased PD risk | [65, 66] |
| **Occupational Role** | Higher education correlates with less physically demanding jobs, potentially increasing PD risk | Sedentary occupations may contribute to neurodegeneration | [71] |
| **Cognitive Reserve** | Greater cognitive reserve may modulate PD symptoms and delay progression | Enhanced cognitive performance may obscure early PD symptoms | [72] |

## 3.3 Environmental and Occupational Factors

PD risk and progression are influenced by various environmental, occupational, socioeconomic, and geographic factors. These factors interact through complex mechanisms that include oxidative stress, neuroinflammation, genetic predisposition, and disrupted cellular processes (Table 4).

**Exposure to Herbicides and Pesticides**

Exposure to herbicides and pesticides, such as paraquat, is strongly associated with an increased risk of PD [73, 74]. Agricultural workers and individuals living in rural areas face heightened exposure through direct contact or environmental contamination. These chemicals induce oxidative stress and mitochondrial dysfunction, critical processes in dopaminergic neuron degeneration, and have been linked to genetic alterations that elevate susceptibility to PD [75].

**Heavy Metal Exposure**

Heavy metals, including mercury, manganese, and lead, disrupt redox homeostasis, leading to excessive reactive oxygen species (ROS) production and impaired antioxidant defense systems [76]. This oxidative stress contributes to mitochondrial dysfunction and the aggregation of alpha-synuclein, forming Lewy bodies—a hallmark of PD [77]. Metals also trigger neuroinflammation through the activation of inflammasomes and proinflammatory cytokines, exacerbating neuronal loss [78, 79]. Additionally, heavy metals cross the blood-brain barrier, accumulate in the brain, and affect gut-brain communication via microbiome disruptions, further amplifying PD risk [79, 80].

**Industrial Chemicals**

Occupational exposure to industrial solvents, such as trichloroethylene, is associated with PD development [76, 81]. These solvents can impair autophagy, lysosomal function, and mitochondrial integrity, leading to neuronal dysfunction [81]. Genetic factors combined with solvent exposure may synergistically increase PD susceptibility.

**Urban and Rural Living**

Urban living, characterized by higher air pollution levels, poses a significant risk for PD [82]. Pollutants contribute to neurotoxicity, systemic inflammation, and genetic interactions that predispose individuals to PD [6, 83, 84]. Particulate matter in polluted air has been shown to directly damage neurons and exacerbate oxidative stress, highlighting the importance of mitigating pollution exposure in urban areas. In contrast, people residing in rural areas as well as those working on farms, in addition to chemical exposures (e.g., pesticides, herbicides), are more prone to head injuries [85], infections [86], soil-borne pathogens[86], and unwell-water consumption—factors that may be causally connected to PD [87].

**Occupational and Workplace Factors**

Certain occupations, particularly in agriculture [88], mining, and industries involving heavy metals [89] or solvents (such as trichloroethylene) [81], elevate PD risk [88]. Agricultural workers face exposure to pesticides and herbicides, while miners encounter high levels of metals like manganese and lead. Professions requiring high cognitive engagement, such as healthcare [90] and teaching [91], may also be linked to PD, potentially due to a combination of environmental exposures and genetic predispositions that affect key cellular processes [92]. Individuals who engage in contact sports, such as boxing, American football, and soccer, are at a heightened risk of experiencing head trauma, particularly repeated head trauma, which increases their susceptibility to developing neurodegenerative disorders like Alzheimer's disease and PD [93].

**Socioeconomic Status**

Lower socioeconomic status (SES), often defined by income and education levels, correlates with higher PD prevalence. Limited access to healthcare among lower SES groups can delay diagnosis and treatment, accelerating disease progression [94, 95]. Conversely, higher SES is associated with lifestyle factors, such as physical activity and dietary antioxidant intake, which reduce PD risk. Environmental exposures, such as pollutants or occupational hazards, are often more prevalent in lower SES populations, further contributing to PD onset and severity.

**Table 4:** Overview of Environmental, Occupational, Socioeconomic, and Geographic Risk Factors in PD.

| Factor | Key Observations | Mechanisms | References |
|---|---|---|---|
| **Herbicides/Pesticides** | Exposure linked to PD risk, particularly among agricultural workers and rural residents | Oxidative stress; mitochondrial dysfunction; genetic alterations | [75] |
| **Heavy Metals** | Exposure to mercury, manganese, lead increases PD risk | Disruption of redox homeostasis; alpha-synuclein aggregation; neuroinflammation; microbiome disruption | [77, 78] |
| **Industrial Chemicals** | Solvents like trichloroethylene associated with PD | Impair autophagy, lysosomal function, and mitochondrial integrity | [92] |
| **Urban Living** | Air pollution contributes to PD risk | Neurotoxicity; systemic inflammation; oxidative stress | [83, 84] |
| **Rural Living** | Increased risk through unwell-water consumption, soil pathogens, head injuries | Environmental exposures; infections; lifestyle factors | [59, 60, 61] |
| **Occupational Roles** | Agriculture, mining, and industries involving solvents and heavy metals linked to PD risk | Environmental toxins; genetic susceptibility; neurotoxic exposures | [71, 79, 90, 91] |
| **Socioeconomic Status** | Lower SES correlates with higher PD prevalence | Limited healthcare access; higher exposure to environmental risk factors; protective lifestyle factors in higher SES groups | [94, 95] |

### 3.4 Lifestyle Factors

Lifestyle factors significantly influence the development of PD [96]. Factors such as physical activity, diet, smoking, and alcohol consumption have been shown to affect the risk of developing PD (Table 5). A thorough understanding of these lifestyle factors is essential for the prevention, early detection, and management of PD.

**Body Mass Index (BMI)**

The relationship between BMI and PD remains contentious. Some studies report no significant link between obesity or prolonged sedentary behavior and PD risk [97]. Conversely, other research suggests that abdominal obesity and increased waist circumference may elevate PD risk [98, 99]. These effects may be mediated by elevated inflammatory cytokines, oxidative stress, and obesity-related metabolic disturbances like insulin resistance and dyslipidemia, which contribute to mitochondrial dysfunction and neuronal damage in PD [100]. Additionally, obesity-induced changes in the gut microbiome may exacerbate inflammation and oxidative stress, further influencing PD progression. Unintentional weight loss, however, is a common feature of PD, affecting approximately 60% of patients, with an average loss of 3-6 kg during disease progression [9]. This weight loss correlates with poor prognosis and diminished quality of life [9, 101].

**Alcohol Consumption**

The relationship between alcohol consumption and PD risk remains inconsistent across studies. While some research has not identified any significant link between alcohol and PD [102], other studies suggest a protective effect of moderate alcohol consumption against motor function decline [103, 104]. Acute alcohol intake may temporarily alleviate motor symptoms, such as tremors and bradykinesia, by increasing dopamine release in specific brain regions. However, heavy and prolonged alcohol use can lead to persistent reductions in dopamine levels, exacerbating PD symptoms over time [105]. Neuroimaging studies reveal that alcohol intake is negatively associated with brain structure and connectivity changes, potentially accelerating PD progression [106]. Additionally, alcohol can interfere with the efficacy of medications like levodopa, a precursor of dopamine, by impairing its absorption and effectiveness. This can intensify tremors and other motor symptoms. Furthermore, alcohol consumption disrupts sleep patterns, compounding existing sleep disturbances commonly experienced by individuals with PD.

**Dietary Habits and Nutrition**

The consumption of dairy products has been associated with an increased risk of developing PD [107], particularly in men [108, 109]. Potential contributing factors include milk proteins such as casein and lactalbumin, which may lower serum urate levels. Since urate has protective effects against PD, this reduction could increase vulnerability [108]. Additionally, contaminants like pesticides in dairy products may play a role [110]. While these associations do not establish causality, they suggest a potential risk link. Diets high in pro-inflammatory foods, such as red meat, refined sugars, and trans fats, may promote systemic inflammation, further increasing PD risk. Similarly, carbonated beverages and cold cuts have also been linked to a higher risk of PD [111]. Nutrient deficiencies, including omega-3 fatty acids, vitamin D, B vitamins, and coenzyme Q10, can impair neuronal health and mitochondrial function, potentially accelerating neurodegeneration and worsening PD symptoms[112]. Moreover, high-protein diets may interfere with the absorption of levodopa, the primary medication for managing PD symptoms.

**Sedentary Lifestyle**

Physical inactivity and a sedentary lifestyle are associated with both the development and progression of PD. Sedentary behavior worsens non-motor symptoms such as cognitive impairment, depression, and poor sleep quality, significantly reducing the quality of life for individuals with PD [96, 113]. Physical inactivity is also correlated with increased depression and cognitive decline in PD patients, suggesting that reducing sedentary time could help alleviate these symptoms. Furthermore, dysregulated genes identified in PD patients overlap with those affected by sedentary behavior, particularly in cellular pathways linked to PD progression. Network-based research underscores the impact of lifestyle factors on significant genetic and molecular pathways, highlighting potential therapeutic targets for slowing PD progression.

**Table 5:** Summary of Lifestyle Risk Factors in PD.

| Category | Key Observations | Mechanisms | References |
|---|---|---|---|
| **Body Mass Index** | Conflicting evidence on association; unintentional weight loss common in PD | Abdominal obesity may elevate risk; weight loss correlates with poor prognosis | [80-84] |
| **Dairy Consumption** | Linked to a higher risk of PD, particularly in men | Decreases serum urate levels; potential contaminants like pesticides | [70-73] |
| **Alcohol Consumption** | Mixed evidence: moderate consumption may reduce PD risk; heavy use may exacerbate progression | Dopamine release (acute use); prolonged use depletes dopamine; interferes with PD medication; impacts brain structure/connectivity | [75-79] |
| **Sedentary Lifestyle** | Associated with worse non-motor symptoms (e.g., cognitive decline, depression) | Dysregulated genes linked to cellular pathways; overlaps with molecular pathways influencing PD progression | [96, 113] |

### 3.5 Metabolic and Systemic Factors

Metabolic and systemic factors can worsen neurodegeneration by promoting inflammation, oxidative stress, and mitochondrial dysfunction, accelerating PD progression (Table 6). Understanding these factors is crucial for developing strategies to manage and slow the disease.

**Diabetes**

Type 2 diabetes mellitus (T2DM) significantly impacts PD development and progression [114-116]. T2DM exacerbates motor symptoms (e.g., instability and mobility issues), non-motor symptoms, cognitive impairments (e.g., slower thinking and attention deficits), and accelerates disease progression [117-120]. The neurovascular burden associated with T2DM, such as white matter hyperintensities, contributes to these effects [121, 122]. T2DM is thus considered an independent factor negatively influencing PD outcomes [122, 123].

**Hypertension**

Hypertension may increase PD risk due to hypertensive vasculopathy in regions such as the basal ganglia, brain stem, and thalamus [124, 125]. It also contributes to cardiovascular diseases and cerebral small vessel diseases, leading to white matter hyperintensities that may exacerbate PD progression [126-128]. Early diagnosis and management of hypertension could mitigate PD risk and progression.

**Cholesterol and Triglycerides**

PD patients often exhibit lower levels of cholesterol and triglycerides compared to healthy controls [129, 130]. Studies have reported consistently reduced total cholesterol, LDL-C, HDL-C, and triglycerides in PD patients up to 20 years before diagnosis [131]. This reduction is associated with increased oxidative stress and alpha-synuclein aggregation in dopaminergic neurons [132]. High plasma LDL-C levels and total cholesterol have been linked to slower PD progression and improved motor function [133, 134], while low HDL-C levels are associated with higher PD risk [135]. However, findings on lipid levels remain inconsistent, likely due to confounding factors such as obesity, diabetes, lifestyle, and genetics [134] [136] [137, 138].

**Table 6:** Overview of Metabolic and Physiological Factors Contributing to Harm in PD.

| Category | Key Observations | Mechanisms | References |
|---|---|---|---|
| **Diabetes** | T2DM linked to more severe motor/non-motor symptoms and accelerated PD progression | Neurovascular burden; white matter hyperintensities; cognitive decline | [85-95] |
| **Hypertension** | May elevate PD risk and worsen progression | Hypertensive vasculopathy; cerebral small vessel disease; white matter hyperintensities | [96-101] |
| **Cholesterol/Triglycerides** | Lower levels associated with higher PD risk; high LDL-C and total cholesterol linked to slower progression | Oxidative stress; alpha-synuclein aggregation; protective effects of LDL-C | [102-112] |

LDL-C: Low-Density Lipoprotein Cholesterol; HDL-C: High-Density Lipoprotein Cholesterol; T2DM: Type 2 Diabetes Mellitus.

## 3.6 Inflammation, Immunity, and the Gut-Brain Axis

Inflammatory factors, immune dysfunction, and gut-brain axis dysbiosis contribute to neurodegeneration and progression in PD [86]. Chronic inflammation and microglial activation release harmful cytokines and oxidative molecules, while dysbiosis disrupts gut function, triggering brain inflammation and neuronal damage. These factors may influence PD onset, as early symptoms often precede central nervous system changes. A summary of their impact on PD is in Table 7. Understanding these mechanisms is key to developing therapies to slow disease progression.

**Viral Infections**

Viral infections play a significant role in the development and progression of PD by contributing to chronic neuroinflammation. This neuroinflammation is triggered when viruses activate the immune response, particularly through the activation of microglia, which release pro-inflammatory cytokines and chemokines. These molecules can cross the blood-brain barrier and cause neuronal cell death [139, 140]. Furthermore, certain viruses, such as the influenza and Epstein-Barr viruses, have been found to promote the aggregation of alpha-synuclein, a protein critical to the pathology of PD [141]. In addition, the activation of Toll-like receptors (TLRs), particularly TLR4, by viral proteins leads to heightened immune responses and further neuroinflammation, ultimately causing neuronal damage [142]. Viral RNA can also trigger microglial activation through receptors like Mac-1, resulting in the production of reactive oxygen species that contribute to neuronal injury [143]. Epidemiological studies have identified specific viruses, including cytomegalovirus (CMV), Epstein-Barr virus (EBV), and hepatitis C virus, as factors that may induce chronic inflammation and immune dysregulation, potentially increasing the risk of PD development [139].

**Neuroinflammation and Chronic Inflammatory Diseases**

Inflammation is a normal immune response designed to protect the body, but when it becomes chronic, it can lead to tissue damage and contribute to the development of diseases like PD. Chronic inflammation in PD is driven by cytokines, reactive oxygen species, and chemokines that are produced by various immune cells, including endothelial cells, microglia, astrocytes, and peripheral immune cells. Inflammation can manifest in two forms: acute, which has a rapid onset and short duration with localized symptoms, and chronic, which persists over time and leads to tissue damage and an increased risk of disease. In PD, markers of inflammation such as CRP, TNF-α, and IL-6 have been linked to the disease and may serve as potential biomarkers and therapeutic targets [144, 145]. The gut-brain axis, wherein inflammation in the gastrointestinal tract, influenced by gut microbiota, can exacerbate neuroinflammation, further promoting PD progression, is also a key factor [146]. Additionally, genetic factors like LRRK2 mutations and environmental exposures such as toxins or infections play a role in driving neuroinflammation in PD [147].

**Gut Dysbiosis**

Gut dysbiosis, an imbalance of gut microbiota, is increasingly recognized as a critical factor in PD pathogenesis. The gut-brain axis, a bidirectional communication system between the gut and central nervous system, is essential for maintaining neurological health. Dysbiosis can disrupt this axis, leading to neuroinflammation and misfolding of alpha-synuclein, a protein closely associated with PD pathology [147, 148]. Altered gut microbiota can produce pathogenic metabolites that create a pro-inflammatory environment, exacerbating neuroinflammation and promoting PD progression [148, 149]. Several biological pathways connect gut dysbiosis to the progression of PD. Dysbiosis may lead to the overproduction of pro-inflammatory cytokines such as tumor necrosis factor (TNF) and interleukins, which can cross the blood-brain barrier and induce neuroinflammation [150]. Additionally, the gut microbiota influences immune responses through toll-like receptors, and dysregulation of these receptors can contribute to alpha-synuclein pathology [150]. Moreover, the gut-brain axis, which relies on neuroendocrine signals and direct neural pathways, can be disrupted by dysbiosis, exacerbating PD progression [151, 152].

**Chronic Constipation**

Chronic constipation is a prevalent non-motor symptom that often precedes the motor manifestations of PD. This condition is linked to gut dysbiosis and increased intestinal permeability, which may result in systemic inflammation and enteric nervous system activation. These changes can trigger alpha-synuclein misfolding in the gut, potentially propagating to the brain via the vagus nerve, thereby contributing to PD development [149]. The early occurrence of gastrointestinal symptoms, including constipation, supports the hypothesis that PD pathology might originate in the gut years before motor symptoms appear [151].

**Table 7:** Summary of Inflammatory, Microbiome, and Gut-Brain Axis Factors in PD.

| Category | Key Observations | Mechanisms | References |
|---|---|---|---|
| **Viral Infections** | Linked to PD through chronic neuroinflammation and immune dysregulation | Activation of microglia, TLRs; alpha-synuclein aggregation; immune responses to viral proteins | [139, 142] |
| **Neuroinflammation** | Contributes to PD onset and progression | Elevated cytokines (e.g., TNF-α, IL-6); microglial activation; oxidative stress | [144, 153] |
| **Chronic Inflammatory Diseases** | Associated with heightened PD risk and progression | Persistent cytokine production; gut-brain axis inflammation; genetic predispositions | [153] [154] [155] [156] |
| **Gut-Brain Axis** | Bidirectional link between gastrointestinal inflammation and neurodegeneration in PD | Shared genetic factors; inflammatory markers in the gut | [146, 147] |
| **Gut Dysbiosis** | Linked to neuroinflammation and alpha-synuclein misfolding; promotes PD progression. | Pro-inflammatory cytokines; toll-like receptor dysregulation; disrupted gut-brain communication. | [147, 148, 150] |
| **Chronic Constipation** | Common preclinical symptom of PD; linked to gut dysbiosis and intestinal permeability. | Alpha-synuclein misfolding in the gut propagating via the vagus nerve to the brain. | [149, 151, 157] |

## 3.7 Neurological and Psychological Factors

Neurological and psychological factors can increase the risk and progression of PD by affecting neural circuits, neuroinflammation, and promoting oxidative stress, which damage neurons. Impaired dopamine regulation, common in both neurological and psychological disorders, may also contribute to the early onset and worsening of PD symptoms. Table 8 provides a summary of the mechanisms underlying neurological and psychological factors in PD.

**Traumatic Brain Injury**

Traumatic Brain Injury (TBI) is a significant risk factor for PD, with biological mechanisms including inflammation, metabolic dysregulation, and protein accumulation. TBI promotes the upregulation of proteins such as amyloid precursor protein (APP), alpha-synuclein, hyper-phosphorylated Tau, and TAR DNA-binding protein 43, all linked to PD pathology [158]. Acute neuroinflammation and catecholamine dysfunction triggered by TBI further contribute to PD progression [159]. The inflammatory response involves microglia and astrocyte activation, which can become detrimental if prolonged, accelerating neurodegeneration[158]. Notably, U.S. veterans with TBI have a 56% increased risk of PD, with the risk escalating with injury severity [158].

**Sleep Disorders**

REM sleep behavior disorder (RBD) serves as an important early indicator of PD. Studies have demonstrated that individuals with RBD are at an increased risk of developing PD or other synucleinopathies, such as dementia with Lewy bodies [160]. The relationship between RBD and PD is believed to stem from the accumulation of misfolded alpha-synuclein proteins in brain regions responsible for regulating REM sleep, leading to the formation of Lewy bodies, a characteristic feature of PD. RBD is marked by the absence of muscle atonia during REM sleep, causing dream enactment, and often precedes the onset of motor symptoms in PD by several years [161]. Beyond RBD, sleep fragmentation and difficulties with sleep maintenance are frequently observed in PD patients, contributing to the worsening of cognitive decline and other non-motor symptoms [160]. As both a risk factor and an early sign of PD progression, RBD reflects the underlying neurodegenerative changes associated with the disease.

**Loss of Smell (Anosmia)**

Anosmia, or the loss of smell, is an early non-motor symptom PD that often precedes motor symptoms by years or even decades. It is linked to the accumulation of alpha-synuclein in the olfactory bulb and related regions, which are among the first areas affected by PD [162]. This pathology begins in the olfactory bulb and related regions, such as the anterior olfactory nucleus, which are among the first areas affected in PD. Reduced integrity of the substantia nigra, a characteristic feature of PD, has also been observed in individuals with unexplained smell loss, further supporting the connection between anosmia and PD. This suggests that anosmia not only signals PD risk but may also serve as an early marker of the disease [163]. The disruption of olfactory function by alpha-synuclein pathology may contribute to the progression of neurodegeneration to other brain regions.

**Chronic Stress**

Psychological and emotional health significantly influence both the risk and progression of PD. The progression of PD is also closely linked to various biological mechanisms influenced by psychological stress and mood disorders. Stress can accelerate the damage to dopamine-producing neurons, worsen motor symptoms like bradykinesia, motor blocking, and tremors, and reduce the effectiveness of dopaminergic medications [164]. The mesocortical and mesolimbic dopaminergic pathways, which are crucial for mood regulation, are disrupted in PD, contributing to the high prevalence of depression and anxiety in patients [165, 166]. Chronic stress has been identified as a significant risk factor for PD, as it exacerbates the loss of dopamine-producing neurons. Animal models suggest that prolonged stress accelerates dopaminergic cell death, underscoring the potential benefits of stress-reduction interventions [167, 168].

**Depression and Anxiety**

Mental health conditions such as depression and anxiety can serve as early indicators during the prodromal phase of PD, though their presence does not necessarily predict the development of the disease. Research shows that individuals under 50 with mental health disorders, including schizophrenia, bipolar disorder, insomnia, depression, and anxiety, are at a heightened risk of PD, particularly when other environmental or contributory factors are present [169, 170]. Depression and anxiety, prevalent non-motor symptoms of PD, may also precede the onset of motor symptoms and are increasingly recognized as integral to the disease's pathophysiology. The biological mechanisms linking stress, depression, and anxiety to PD include the activation of the hypothalamus-pituitary-adrenal (HPA) axis and the sympathetic nervous system (SNS), elevated levels of glucocorticoids and catecholamines leading to neuronal injury, increased proinflammatory cytokines contributing to neurodegeneration, and stress-induced vulnerability of dopamine neurons via glucocorticoid receptor activity. Chronic stress and mood disorders not only exacerbate both motor and non-motor symptoms but also synergize with other factors to increase neuronal vulnerability, potentially accelerating the progression of PD. Therefore, addressing stress and mood disorders in PD patients is crucial for slowing the disease's progression and improving their quality of life [171].

**Table 8:** Overview of Neurological and Psychological Risk Factors PD.

| Category | Key Observations | Mechanisms | References |
|---|---|---|---|
| **Traumatic Brain Injury** | Increases PD risk by 56%; severity dependent. | Inflammation, protein accumulation (APP, alpha-synuclein, etc.), microglial activation. | [158, 159] |
| **Sleep Disorders** | REM sleep behavior disorder often precedes motor symptoms by years. | Degeneration of brainstem nuclei; alpha-synuclein accumulation forming Lewy bodies. | [160, 161] |
| **Loss of Smell (Anosmia)** | Early non-motor symptom, precedes motor signs. | Alpha-synuclein accumulation in olfactory pathways. | [162, 163] |
| **Chronic Stress** | Increases neuronal vulnerability; exacerbates motor and non-motor symptoms. | Disrupts dopaminergic pathways; increases proinflammatory cytokines; accelerates cell death. | [164, 167] |
| **Depression and Anxiety** | Common non-motor symptoms; may precede motor symptom onset. | Activates HPA axis and SNS; increases glucocorticoids and catecholamines; induces neuroinflammation. | [165, 171] |

### 3.8 Hormonal and Other Health Conditions

Hormonal factors influence the onset, progression, and severity of PD. Imbalances in hormones such as thyroid and cortisol can affect neuronal health and exacerbate symptoms. Understanding these factors is essential for developing gender-specific therapies and improving disease management. A summary of hormonal risk factors in PD is presented in Table 9.

**Estrogen and Its Neuroprotective Role**

Estrogen plays a significant role in the risk and progression of PD by influencing key processes involved in dopaminergic function, such as dopamine synthesis, metabolism, and transport. It is believed to have neuroprotective effects, potentially shielding dopaminergic neurons from degeneration. Animal models support this, showing that estrogen may reduce inflammation by interacting with the brain's renin-angiotensin system and counteract oxidative stress, which can damage neurons. Through these mechanisms, estrogen helps protect the brain's dopaminergic system, potentially slowing PD progression [172].

**Postmenopausal Women and PD Risk**

Postmenopausal women, particularly those not undergoing hormone replacement therapy (HRT), face an increased risk of PD. This heightened risk is linked to reduced lifetime exposure to endogenous estrogens, whether due to early menopause or a shorter fertile lifespan. The absence of estrogen supplementation through HRT after menopause further exacerbates the risk [173, 174].

**Hormone Replacement Therapy**

The link between hormone replacement therapy (HRT) and PD risk is multifaceted. While estrogen alone may offer protective effects, long-term use of combined estrogen-progesterone therapy appears to increase the risk of developing PD [172, 175]. Estrogen replacement therapy (ERT) can help alleviate symptoms and slow the progression of PD in women, especially in the early stages before L-dopa therapy begins [176]. The lower incidence of PD in women compared to men may be partly attributed to the protective influence of endogenous estrogen. Estrogen is thought to protect against neurodegeneration by reducing oxidative stress, regulating dopamine metabolism, and improving mitochondrial function, which may help preserve dopaminergic neurons in the substantia nigra, a region progressively affected by PD [177].

**Hyperuricemia**

Uric acid, with its antioxidant properties, helps reduce oxidative stress, a key factor in neurodegeneration in PD. The relationship between uric acid levels and PD risk is complex. Elevated uric acid levels are associated with a reduced risk of PD due to their neuroprotective effects. However, hyperuricemia is also linked to cognitive impairment and conditions like gout, which may exacerbate neurodegeneration [178]. Additionally, co-existing conditions such as cerebrovascular injury can amplify the effects of hyperuricemia, increasing the risk of cognitive decline and accelerating PD progression [179].

**Table 9:** Overview of Hormonal and Other Health Condition Risk Factors in PD.

| Category | Key Observations | Mechanisms | References |
|---|---|---|---|
| **Estrogen Levels** | May protect dopaminergic neurons; reduced levels post-menopause linked to higher PD risk. | Modulates dopamine pathways; reduces inflammation and oxidative stress. | [172-174] |
| **Postmenopausal Women** | Higher PD risk linked to early menopause, shorter fertile life, and lack of HRT. | Reduced lifetime estrogen exposure affects neuroprotection. | [173, 174] |
| **HRT** | Mixed effects: long-term combined estrogen-progesterone therapy may increase risk. | Duration- and regimen-specific impact on estrogen's neuroprotective role. | [172, 175] |
| **Hyperuricemia** | Paradoxical effects: antioxidant properties may protect but linked to cognitive decline and gout. | Reduces oxidative stress but compounds neurodegeneration with co-existing conditions. | [178, 179] |

**3.9 Medications**

Certain antibiotics, calcium channel blockers (treating high blood pressure and heart conditions), certain anti-diabetic medications (such as rosiglitazone and pioglitazone), and statins (treating high cholesterol levels) are possible associations with an increased risk of PD [4, 180, 181]. A summary of medications associated with an increased risk of Parkinson's disease is presented in Table 10.

**β2-Adrenoceptor Antagonists (e.g., Propranolol)**

Propranolol, a β2-adrenoceptor antagonist, has been associated with an increased risk of PD. β2-adrenoceptors play a critical role in regulating metabolic functions and dopaminergic activity, both of which are essential for neuroprotection. Chronic use of propranolol may contribute to the degeneration of dopaminergic neurons in the substantia nigra, a hallmark of PD. The β2-adrenoceptor blockade can exacerbate motor symptoms and accelerate PD progression by reducing dopaminergic signaling. Additionally, the modulation of adrenergic activity may disrupt the already compromised balance of neurotransmitters, further impairing dopaminergic systems [182].

**Diabetes Medications**

Metformin, a widely used diabetes medication, has been linked to an increased risk of PD due to its influence on various physiological processes. It can contribute to insulin dysregulation, leading to neuroinflammation and mitochondrial dysfunction, and may also alter synaptic plasticity, implicating neurodegeneration pathways that are common to both diabetes and PD. In contrast, other diabetes treatments, such as glucagon-like peptide-1 agonists (GLP1a), may offer protective effects against PD progression by modulating insulin signaling [183, 184].

**Antibiotics**

Macrolides and lincosamides, classes of antibiotics, have been associated with an increased risk of PD, likely due to their disruptive effects on gut microbiota. Alterations in microbial composition may exacerbate neuroinflammation and promote the aggregation of α-synuclein, a protein implicated in PD pathogenesis. In contrast, emerging therapies such as prasinezumab, a monoclonal antibody targeting aggregated α-synuclein, show potential in slowing motor progression in rapidly advancing PD cases [185, 186].

**Psychological Medications**

Psychotropic medications, including antidepressants and anxiolytics, have been linked to an increased risk of PD in older populations. These drugs can impact dopaminergic systems, with antipsychotics in particular blocking dopamine receptors, which may increase vulnerability to PD [187].

**Cancer Medications**

Although direct evidence is limited, certain chemotherapeutic agents and cancer medications may elevate the risk of PD through mechanisms such as inducing oxidative stress, which generates reactive oxygen species (ROS) and exacerbates neuronal damage, and causing mitochondrial dysfunction, which impairs cellular energy production—a known contributor to PD pathology [188].

**Epigenetic Modifications**

Certain medications can modify epigenetic marks, thereby influencing gene expression associated with PD. Prolonged use of these drugs may induce changes in DNA methylation, a modification commonly observed in PD patients [189]. Epigenetic alterations, including DNA methylation, histone modifications, and hydroxy methylation, are integral to the molecular mechanisms underlying PD, facilitating the interaction between genetic and environmental factors that contribute to disease onset and progression [44].

**Table 10:** Summary of Medications Associated with Increased Risk of PD.

| Factor | Key Observations | Mechanisms | References |
| --- | --- | --- | --- |
| β2-Adrenoceptor Antagonists | Linked to increased PD risk, particularly with chronic use (e.g., propranolol). | Disrupts dopaminergic activity and metabolic functions; accelerates Lewy body formation. | [182] |
| Diabetes Medications | Mixed effects: metformin may increase risk, while GLP1a treatments may be protective. | Insulin dysregulation; mitochondrial dysfunction; neuroinflammation. | [183, 184] |
| Antibiotics | Macrolides and lincosamides associated with higher PD risk. | Disrupt gut microbiota; exacerbate neuroinflammation and α-synuclein aggregation. | [185, 186] |
| Psychotropic Medications | Increased PD risk observed in older populations using antidepressants and anxiolytics. | Blocks dopamine receptors, affecting dopaminergic systems. | [187] |
| Cancer Medications | Potential increase in PD risk with certain chemotherapy agents. | Induces oxidative stress and mitochondrial dysfunction; affects the immune system. | [188] |
| Epigenetic Modifications | Long-term medication use may influence PD through altered DNA methylation. | Changes in epigenetic marks affecting gene expression. | [44, 189] |

## 3.10 Comorbid Conditions

Other neurodegenerative or chronic diseases may increase risk or worsen progression.

## 4. Prevent Factors in PD

Preventive factors in PD focus on strategies aimed at reducing the risk or delaying the onset of this progressive neurodegenerative disorder (Fig. 3). Research indicates that lifestyle choices, nutrition, and occupational factors may play a role in lowering the risk of developing PD. Identifying and understanding these factors can guide the development of strategies that not only delay PD onset but also potentially reduce its overall prevalence. By incorporating these preventive measures, individuals may lower their risk of developing PD or slow its progression.

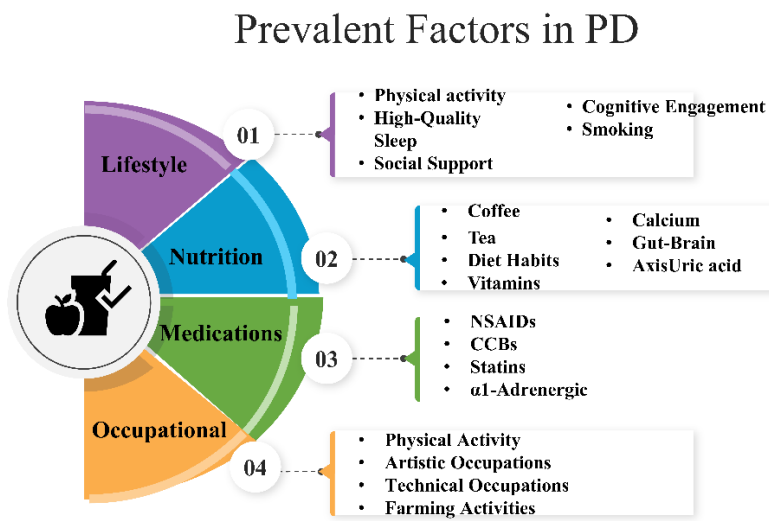

**Figure 3**: Important Preventive Factors in PD.

### 4.1 Lifestyle Factors

Lifestyle factors play a crucial role in both preventing the onset of PD and slowing its progression. Research indicates that specific lifestyle habits can have protective effects on brain health, potentially reducing the risk of developing PD, delaying its onset, and mitigating its progression and symptoms. A summary of lifestyle factors that may help prevent PD is provided in Table 11.

**Physical activity**

Engaging in the highest level of physical activity and exercise has the advantage of decreasing the incidence of PD by up to 21% [190]. It is possible that the onset of PD could be delayed by a significant amount of physical activity, as it may slow down the disease's pathological processes [191], particularly in individuals who are more physically active during their middle years [192]. During the prodromal phase of PD, individuals may experience nonmotor symptoms such as sleep disorders and constipation, along with subtle motor signs like rigidity, balance impairment, and tremor, leading to a likely decrease in physical activity during this period [191]. Patients in the early stages of PD who have consistently engaged in physical activity and exercise over a long period of time may notice a slower decline in gait stability, activities of daily living, and processing speed [193]. Additionally, participating in physical activity and exercise can improve non-motor symptoms like depression, apathy, and postural instability, while also reducing motor symptoms by reducing inflammation in individuals with PD [194]. Consequently, maintaining a high level of physical activity and exercise could stimulate interest in potential therapeutic targets for PD, ultimately contributing to an improved quality of life for individuals with the condition. Vigorous physical activity and reduced sitting time are noted for their neuroprotective effects, including improved gait stability, daily living activities, and processing speed [68]. Exercise has been shown to alleviate non-motor symptoms such as depression, apathy, and postural instability while reducing motor symptoms through anti-inflammatory mechanisms [69]. Engaging in regular physical activity may also delay PD onset and contribute to a better quality of life.

**High-Quality Sleep**

High-quality sleep plays a crucial role in reducing the risk of PD through several interconnected biological mechanisms. One key factor is the promotion of efficient protein clearance, particularly through the glymphatic system, which becomes more active during sleep, clearing toxic proteins like α-synuclein that contribute to PD. Sleep also regulates neuroinflammation by reducing the activation of immune cells, such as microglia and astrocytes, which can exacerbate neural injury [195]. Additionally, good sleep helps mitigate oxidative stress by reducing the production of reactive oxygen species and supporting the body's antioxidant systems, both of which are critical in protecting brain cells from damage. Furthermore, maintaining regular circadian rhythms through high-quality sleep is vital for dopamine regulation, which is essential for PD prevention [196]. Disruptions in circadian rhythms can reduce dopamine levels, increasing PD risk. Sleep also improves sleep architecture, reducing fragmentation and enhancing restorative sleep, which supports overall brain health [196]. These combined mechanisms highlight the neuroprotective role of sleep, suggesting that interventions to improve sleep quality may help delay or reduce the onset of PD [197].

**Social Support**

Social support plays a crucial role in reducing the risk of PD and impacts both psychological and physical health. Studies show that strong social networks are linked to better mental health, reducing depression, anxiety, and stress—key risk factors for PD [198]. These connections may lower the likelihood of developing the disease. Social support also helps slow PD progression through several mechanisms. It encourages physical activity, which has been shown to improve motor and cognitive functions in PD patients by enhancing cerebral mechanisms and cardiovascular fitness [199, 200]. Additionally, the psychological benefits of social support, such as increased positivity and reduced distress, can improve disease management and treatment adherence [198]. Furthermore, access to information through social networks provides patients with strategies, like dietary changes, that may help slow progression, such as modifications that influence the gut microbiome [201].

**Cognitive Engagement**

Cognitive engagement, such as participating in mentally stimulating activities, may help reduce the risk of developing PD and its associated cognitive decline. One mechanism through which cognitive engagement may protect against PD is by enhancing neuroplasticity, the brain's ability to reorganize and form new neural connections. This process, stimulated by engaging in complex tasks, may build cognitive reserve, potentially delaying the onset of cognitive symptoms in PD [202]. Additionally, cognitive engagement may support neurotransmitter systems crucial in PD, such as dopamine, serotonin, noradrenaline, and acetylcholine, which are implicated in cognitive processes and affected in PD. Engaging in mental activities may help maintain these systems, potentially slowing cognitive decline [203]. Moreover, cognitive engagement may influence mitochondrial function, reducing oxidative stress and improving energy metabolism in neurons, which could slow cognitive deterioration [204]. It may also interact with genetic factors, such as variations in the APOE and GBA genes, which are linked to cognitive decline in PD, potentially mitigating their effects [205]. Cognitive engagement can also help slow PD progression by building cognitive reserve, allowing individuals to compensate for neural damage and delay cognitive symptoms [206]. Furthermore, cognitive activities are often associated with increased physical activity, which has been shown to enhance brain function and structure, leading to improved cognitive outcomes in PD patients [207]. Cognitive engagement may also help reduce anxiety, indirectly supporting cognitive health and slowing disease progression [207].

**Smoking**

It is undeniable that smoking tobacco cigarettes is detrimental to our health. Several studies have shown that cigarette smoking may reduce the risk of developing PD [208], with smokers found to be only half as likely as non-smokers to develop PD [209]. A significant inverse dose-response correlation has been discovered in relation to PD risk: the greater the duration and frequency of smoking among individuals, the lower their chances of developing PD [6]. Consequently, smoking has been linked to a later age of onset of PD [2]. The potential protective effect of smoking on PD may be due to the presence of nicotine in tobacco smoke, as it may trigger the release of dopamine, a neurotransmitter associated with PD [208]. Moreover, nicotine has the ability to block the formation of toxins by directly impacting enzyme functions, such as MAO-B [210]. Cigarette smoking has also been observed to diminish the aggregation of alpha-synuclein in cell cultures, suggesting a potential slowdown in the development of PD[211]. Although cigarette smoking may have some protective effects in PD, it was found to be directly linked to increased issues with drooling, swallowing, and freezing in patients with PD [7]. PD smokers were more likely to experience non-motor symptoms such as unexplained pains, feelings of sadness, and memory problems [7]. Moreover, while there may not be a strong connection between smoking and the mortality of PD, it is clear that individuals with PD who smoke are at a higher risk of developing smoking-related cancers like lung cancer, cardiovascular disease, and respiratory issues [212].

Table 11: Overview of Lifestyle Factors Potentially Preventing PD.

| Factor | Key Observations | Mechanisms | References |
|---|---|---|---|
| **Physical Activity** | Reduces PD incidence by up to 21%, delays onset, and improves symptoms. | Enhances neuroprotection, reduces inflammation, and improves gait and quality of life. | [190, 192, 193] |
| **High-Quality Sleep** | Reduces PD risk by promoting protein clearance and dopamine regulation. | Clears toxic proteins, reduces neuroinflammation, and supports brain health. | [195-197, 213] |

| Social Support | Reduces PD risk, supports mental health, and encourages physical activity. | Reduces stress, improves motor/cognitive functions, and supports disease management. | [198-200] |
| Cognitive Engagement | Mental activities reduce PD risk and cognitive decline by enhancing neuroplasticity. | Boosts brain function, maintains neurotransmitters, and supports mitochondrial health. | [202-205] |
| Smoking | Smokers are less likely to develop PD, but it worsens non-motor symptoms and other health risks. | Nicotine may trigger dopamine release and reduce alpha-synuclein aggregation but increases other health risks. | [69, 208, 209, 211, 212, 214] |

## 4.2 Dietary and Nutritional Factors

Dietary and nutritional factors are crucial in the context of PD as they can influence both the risk of developing the disease and the management of its symptoms. Proper nutrition supports brain health and neuroprotection, reduces inflammation, promotes gut health, and maintains overall well-being—all of which are vital in managing a neurodegenerative condition like PD. Additionally, the interactions between medications and dietary factors are an important consideration in treatment. A summary of dietary and nutritional factors that may have a preventive influence on PD is provided in Table 12.

**Coffee**

Drinking coffee may help prevent the onset of PD, by ~30% [214-216]. Additionally, consuming coffee is linked to a decreased likelihood of developing PD as well as slowing its advancement [217]. A recent study has reported a strong correlation between the duration of coffee consumption and the age of onset of PD, as well as a relatively low association between the number of cups of coffee and the age of onset of PD [214]. The potential explanation for the protective effect of coffee against PD is attributed to caffeine. This is supported by the fact that black tea, which contains a lower amount of caffeine, demonstrated a more moderate association with the onset age of PD [214]. There is a paradox regarding the effects of caffeine on motor and non-motor symptoms among PD patients. On one hand, research on drug-naive, early-stage PD patients has shown that the severity of non-motor symptoms related to mood and cognition has an inverse relationship with coffee consumption [218]. Additionally, some research has indicated that individuals who consume coffee exhibit lower tremor scores compared to those who do not, with a correlation observed between the severity of tremors and the amount of coffee consumed [219]. Also, Coffee was found to have a protective effect against motor function decline, mortality, and cognitive decline in PD patients [105]. On the other hand, some studies have shown that there is no meaningful connection between the dosage and duration of coffee consumption and the presence of motor or non-motor symptoms in early-stage PD patients [7]. There are multiple reasons why coffee may offer protection against the development of PD. One of the reasons is that a component of coffee acts as an adenosine A2A receptor antagonist, leading to decreasing α-synuclein aggregation in SynT-Synphilin-1 neuroglioma cells [220]. Additionally, coffee consumption not only exhibits anti-inflammatory properties but also mitigates the risk of chronic diseases like diabetes, obesity, and cardiovascular disease, all of which indirectly influence the development and progression of PD [221].

**Tea**

Multiple studies and meta-analyses have consistently shown an inverse relationship between tea consumption and the risk of developing PD. Increased tea consumption is associated with a reduced risk of developing PD [222], with a potential decrease in PD risk by 26% [223]. While most studies focus on the preventive effects of tea consumption, emerging evidence suggests that tea may also have the potential to slow down PD progression. A study highlighted the role of tea drinking in modifying the progression of idiopathic REM sleep behavior disorder (iRBD) to α-synucleinopathies, including PD [224]. The study found that tea drinking was associated with a decreased risk of phenoconversion from iRBD to PD, suggesting a protective role of tea in the early stages of neurodegenerative disease progression [224]. Tea is rich in polyphenols, especially catechins like EGCG, which possess significant neuroprotective properties. These compounds reduce oxidative stress, a major factor in PD pathogenesis, by scavenging free radicals and enhancing antioxidant defenses [223]. Polyphenols also modulate signaling pathways and metal chelation, which are crucial in reducing neuronal damage. Studies on black tea have consistently shown a significant inverse association between black tea consumption and PD risk. The Singapore Chinese Health Study found that black tea consumption was associated with a reduced risk of PD, independent of caffeine intake or tobacco smoking [225]. This protective effect is thought to be due to the presence of thearubigins and theaflavins, which are formed during the oxidation of catechins in black tea. While green tea is rich in polyphenols, particularly catechins like epigallocatechin gallate (EGCG), its association with reduced PD risk in human populations remains inconclusive [225]. Some studies have not shown a significant association between green tea consumption and reduced PD risk [225]. However, green tea polyphenols (GTPs) have demonstrated protective effects against dopaminergic neuron degeneration in preclinical models, suggesting potential neuroprotective properties [225]. A study in Japan found that the intake of Japanese and Chinese teas was significantly inversely associated with the risk of PD, supporting the notion that various types of tea can be beneficial [226]. Theanine, an amino acid found in tea, can inhibit glutamate receptors and regulate extracellular glutamine concentrations, providing neuroprotective effects by reducing excitotoxicity [223].

**Diet Habits**

Diet habits are factor that can have both negative and positive impacts on PD. A dietary regimen characterized by elevated consumption of fruits, fish, vegetables, whole grains, nuts, legumes, and poultry, coupled with reduced intake of saturated fats and moderate alcohol consumption, potentially serves as a protective factor against PD [227]. The relationship between nutrition and PD is complex, involving various dietary patterns and specific nutrients that may influence the risk and progression of the disease. The Mediterranean diet, which is rich in fruits, vegetables, nuts, and fish, has been associated with a reduced risk of PD and slower disease progression. This diet is thought to exert its effects through anti-inflammatory and antioxidant mechanisms, which help mitigate oxidative stress and neuroinflammation, key factors in PD pathogenesis [228, 229].

**Vitamins**

longitudinal studies have indicated that a persistent deficiency in vitamin D is associated with higher likelihood of developing PD[230]. The increased risk may be due to the chronic lack of vitamin D, which can result in the depletion of dopaminergic neurons in the substantia nigra region and ultimately contribute to the onset of PD [230]. Moreover, previous studies have shown that there are seasonal and sunlight-related impacts on dopamine transporter (DAT) expression in both individuals without any health issues and those in the early stages of PD [231]. It has been suggested that dietary vitamin E may play a role in protecting against PD, indicating that incorporating Vitamin E-rich foods into one's diet could be an immediate option for reducing the risk of PD [232]. PD patients have been found to have lower levels of vitamin B12 than healthy controls, which could potentially be linked to the onset of early gait instability and neuropathy[233]. Additionally, it was observed that PD patients who did not experience dementia had elevated levels of vitamin B12 than cognitively healthy PD ones, suggesting that that a higher vitamin B12 level might reduce risk of developing dementia in the future [234].

Similarly, adequate intake of vitamins such as B6 and folate may support neurological health and reduce PD risk by lowering homocysteine levels, which are linked to neurodegeneration [235]. Minerals like magnesium and zinc are also important, as they are involved in numerous enzymatic processes that support neuronal function and may help slow PD progression [235].

**Calcium**

Calcium's role in PD is less clear, but it is essential for various cellular processes, including neurotransmitter release and neuronal excitability. Dysregulation of calcium homeostasis can contribute to neuronal death, a hallmark of PD. Therefore, maintaining adequate calcium levels through diet may support neuronal health and potentially slow disease progression [235].

**Gut-Brain Axis**

Emerging research highlights the gut-brain axis as a significant factor in PD. The gut microbiota can influence neuroinflammation and neurodegeneration, suggesting that dietary interventions that promote a healthy gut microbiome may impact PD risk and progression. Diets rich in fiber and probiotics may enhance gut health and, consequently, brain health, offering a potential therapeutic avenue for PD [236].

**Uric acid (urate levels)**

Recent studies highlight the potential of uric acid (UA) in reducing PD risk. A 1 mg/dL increase in serum urate is associated with a 6% reduction in PD risk [237]. As a potent endogenous antioxidant, UA accounts for 60–70% of antioxidant activity in human plasma [238], and its properties are especially relevant in PD, where oxidative stress is a key factor in neuronal damage [239]. UA mitigates oxidative stress by scavenging free radicals and reactive oxygen species (ROS) in neuronal cells. It also inhibits lipid peroxidation, protecting cellular components from damage. Mitochondrial dysfunction is central to PD, and UA's ability to protect mitochondrial integrity may help preserve neuronal function [239]. Additionally, UA may reduce inflammatory damage to neurons by modulating inflammatory pathways activated in PD [239]. It also aids in maintaining calcium homeostasis, a critical factor in neuronal function, as disruptions in calcium signaling are implicated in PD pathogenesis. One proposed mechanism of UA's neuroprotective effect is the activation of Nrf2, a transcription factor that regulates antioxidant proteins, offering protection against oxidative damage in PD. Research comparing UA levels in PD patients and healthy controls shows significantly lower serum and cerebrospinal fluid (CSF) UA levels in PD patients, suggesting factors beyond purine metabolism, such as age, sex, and weight, may influence UA levels [240]. Higher serum urate is associated with a reduced PD risk, supporting UA's potential neuroprotective effects [237]. Despite these findings, some studies report no significant correlation between high UA levels and reduced PD risk, and the role of UA in PD progression remains debated [240]. Additionally, the causality of low UA levels in PD is still under discussion [240].

**Table 12:** Summary of Dietary and Nutritional Factors That May Help Prevent PD.

| Factor | Key Observations | Mechanisms | References |
|---|---|---|---|
| **Coffee** | Coffee may reduce PD risk by ~30% and slow progression. Inconsistent effects on motor/non-motor symptoms in PD patients. | Caffeine acts as an adenosine A2A receptor antagonist, reducing α-synuclein aggregation and inflammation. | [214, 215, 218, 221] |
| **Tea** | Tea consumption is linked to a 26% reduction in PD risk. Black tea may be more protective than green tea. | Polyphenols (catechins, theaflavins) reduce oxidative stress, modulate signaling, and chelate metals. | [222, 223, 225, 226] |
| **Nutrition and Diet** | A Mediterranean diet (fruits, vegetables, fish) is linked to a reduced PD risk and slower progression. | Anti-inflammatory and antioxidant effects help mitigate oxidative stress and neuroinflammation. | [228, 229] |
| **Vitamins** | Vitamin D deficiency increases PD risk; Vitamin E, B12, B6, and folate may help protect against PD and reduce disease progression. | Vitamins support neuronal function, reduce oxidative stress, and lower homocysteine, which is linked to neurodegeneration. | [179, 228, 235] |
| **Calcium** | The role of calcium in PD is unclear, but it is crucial for neuronal processes. Dysregulation may contribute to neuronal death. | Maintaining calcium homeostasis supports neurotransmitter release and neuronal health. | [235] |
| **Gut-Brain Axis** | A healthy gut microbiome may influence PD risk and progression. Diets rich in fiber and probiotics may promote brain health. | Improves gut health, reducing neuroinflammation and supporting brain health through the gut-brain connection. | [236] |
| **Uric Acid (UA)** | Higher uric acid levels may reduce PD risk by 6% for every 1 mg/dL increase. UA acts as an antioxidant, protecting against oxidative stress. | UA scavenges free radicals, protects mitochondrial integrity, and modulates inflammation, supporting neuronal function. | [178, 238, 240] |

## 4-3 Medications and Medical Factors

Medications and medical factors are vital in preventing or slowing PD progression. For example, anti-inflammatory drugs can reduce neuroinflammation, which is associated with PD development, potentially slowing disease progression. Besides, neuroprotective medications help preserve neurons by neutralizing free radicals and supporting mitochondrial function. Table 13 provides an overview of medications and medical factors, along with their respective mechanisms, in the prevention and slowing of PD progression.

**Non-Steroidal Anti-Inflammatory Drugs (NSAIDs)**

NSAIDs have been explored for their potential neuroprotective effects in PD due to their ability to inhibit cyclooxygenase (COX) enzymes, particularly COX-2, which is upregulated in PD [241]. By inhibiting COX-2, NSAIDs reduce the production of pro-inflammatory prostaglandins, potentially mitigating neuroinflammation and slowing neurodegeneration [241]. NSAIDs such as aspirin and ibuprofen may also decrease elevated cytokines like TNF-α and IL-1β, which contribute to neuroinflammation and neuronal death in PD [242]. Aspirin and Ibuprofen have been shown to effectively reduce the risk of PD and postpone the age of onset in Parkinson's Disease by as much as 5 years [214, 243]. It could be that the anti-inflammatory effects of Aspirin and Ibuprofen play a role in decreasing PD risk [244]. Furthermore, aspirin has been proven to boost the production of tyrosine hydroxylase (TH), the enzyme that controls the rate of dopamine (DA) synthesis. This leads to an elevation in the level of DA found in dopaminergic neurons [245]. Additionally, aspirin and ibuprofen have the potential to decrease the likelihood of LRRK2 mutation, indicating that anti-inflammatory medications could serve as beneficial treatments for modifying the progression of LRRK2-PD [246]. Epidemiological studies suggest that regular NSAID use is associated with a reduced risk of developing PD, although results are inconsistent across studies. Experimental models also demonstrate protective effects of NSAIDs against neurotoxin-induced dopaminergic neuron degeneration. However, long-term NSAID use is linked to gastrointestinal and cardiovascular risks, warranting careful consideration in PD treatment.

**Calcium Channel Blockers (CCBs)**

CCBs have been studied for their potential to reduce the risk and progression of PD by modulating calcium ion flows, which are crucial for neurotransmitter release and neuronal health. In dopaminergic neurons, particularly those in the substantia nigra, CCBs help regulate calcium influx, which is critical given the vulnerability of these neurons to calcium-induced stress due to their pacemaking activity [247]. L-type calcium channels, such as those with CaV1.3 subunits, contribute to mitochondrial oxidant stress, a key factor in PD pathogenesis. Recent research also suggests that Cav2.3 channels may play a role in regulating neuronal viability in the substantia nigra, potentially offering neuroprotective benefits [248]. While a systematic review and meta-analysis found that CCB use was associated with a significantly reduced risk of PD (relative risk: 0.78; 95% CI: 0.62–0.99) [249], other studies, such as a large prospective cohort study, found no significant association [250]. Additionally, a clinical trial with isradipine, a CCB, did not show significant effects on slowing early-stage PD progression [249].

**Statins**

Statins, primarily known for their cholesterol-lowering effects, have been investigated for their neuroprotective potential in PD due to their anti-inflammatory, anti-oxidative, and neuroprotective properties. By suppressing proinflammatory molecules and microglial activation, statins help reduce neuroinflammation in PD. These effects are mediated through inhibition of the mevalonate pathway, which is crucial for synthesizing inflammatory mediators. Statins also inhibit oxidative stress, a major factor in PD pathogenesis, potentially protecting dopaminergic neurons from degeneration. Furthermore, statins can act as ligands for PPARα, a regulator of mitochondrial function, which is often compromised in PD [251]. Evidence suggests that statins, particularly simvastatin due to its ability to cross the blood-brain barrier, may slow PD progression by attenuating α-synuclein aggregation [251]. A systematic review and meta-analysis have suggested that atorvastatin may reduce PD risk, attributed to its potent anti-inflammatory and anti-oxidative effects [252]. However, further randomized controlled trials are needed to confirm these findings.

**α1-Adrenergic Antagonists**

α1-Adrenergic antagonists, such as terazosin, doxazosin, and alfuzosin, primarily used to treat benign prostatic hyperplasia, have shown potential neuroprotective effects in PD by enhancing glycolysis and energy metabolism, which are impaired in PD [253]. These drugs activate PGK1, the first ATP-generating enzyme in glycolysis, increasing ATP production and potentially counteracting the energy deficits observed in PD. Moreover, α1-adrenoceptor antagonists may modulate the locus coeruleus, a key noradrenergic center affected in PD, potentially influencing dopamine release and providing neuroprotection. A cohort study involving Danish and U.S. health registries found that users of terazosin, doxazosin, and alfuzosin had a significantly lower risk of developing PD compared to users of tamsulosin, which does not enhance glycolysis. The hazard ratios for developing PD were 0.88 in the Danish cohort and 0.63 in the U.S. cohort, indicating a reduced risk associated with these drugs [253].

**Table 13:** Overview of medications and medical factors that may potentially prevent the risk of PD.

| Factor | Key Observations | Mechanisms | References |
|---|---|---|---|
| **Non-Steroidal Anti-Inflammatory Drugs (NSAIDs)** | Aspirin and ibuprofen may reduce PD risk and postpone the age of onset by as much as 5 years. Regular use is linked to a reduced risk of PD. | Inhibit COX-2 enzymes, reduce pro-inflammatory prostaglandins, decrease elevated cytokines like TNF-α and IL-1β, potentially mitigate neuroinflammation. Aspirin boosts dopamine production. | [241, 242, 245, 246] |
| **Calcium Channel Blockers (CCBs)** | CCBs, like isradipine, may reduce PD risk but have mixed results in slowing progression. | Modulate calcium ion flows to regulate neurotransmitter release, reduce mitochondrial oxidant stress, and protect dopaminergic neurons. | [247-250] |
| **Statins** | Statins, particularly simvastatin, may slow PD progression by attenuating α-synuclein aggregation and reducing neuroinflammation. | Inhibit pro-inflammatory molecules, reduce oxidative stress, and regulate mitochondrial function via PPARα. | [251, 252] |
| **α1-Adrenergic Antagonists** | Terazosin, doxazosin, and alfuzosin show reduced PD risk due to enhanced glycolysis and energy metabolism. | Activate PGK1 enzyme in glycolysis to increase ATP production, modulate the locus coeruleus to influence dopamine release and provide neuroprotection. | [253] |

### 4.4 Occupational Factors

Certain occupational categories, particularly those involving high levels of physical activity, have been shown to significantly reduce the risk of developing PD. Occupations such as engineering, production work (including machine operators and fabricators), metalworking, and construction or extractive work (e.g., mining and oil well drilling) have been associated with lower PD risk [254]. Research has highlighted that jobs requiring significant physical exertion are linked to a decreased risk of PD, consistent with the well-established protective effects of regular physical exercise (Table 14). Such occupations, which demand consistent physical activity, may confer similar neuroprotective benefits as structured exercise regimens. Multiple studies have also demonstrated a strong association between artistic occupations and a reduced risk of PD. For instance, one study indicated that men engaged in artistic professions later in life had a significantly lower risk of developing PD [255]. This protective effect is thought to be related to the dopaminergic activity required for creativity, which may be better preserved in individuals working in artistic fields [255]. A case-control study conducted in Japan found that professional or technical occupations were inversely related to the risk of PD, particularly among men. This suggests that careers requiring high levels of cognitive engagement and technical skills may have a protective effect against the disease [256]. Interestingly, certain farming activities have been linked to a lower risk of PD. A study of French farm managers revealed that activities such as gardening, landscaping, reforestation, small animal farming, and horse-related tasks (e.g., training, dressage, and riding) were associated with reduced PD risk [257]. This finding is particularly notable given that farming has typically been associated with increased PD risk due to pesticide exposure. Occupations that involve physical activity and cognitive engagement may enhance neuroplasticity and promote overall brain health. Such activities are thought to reduce the risk of neurodegenerative diseases like PD by maintaining robust neural networks and fostering neuronal resilience. Physical activity, which is a key component of many protective occupations, has also been linked to improved mitochondrial function and reduced oxidative stress—both of which are crucial for maintaining neuronal health and potentially slowing neurodegenerative processes. Additionally, occupations that involve regular physical activity or cognitive stimulation may increase the production of neurotrophic factors, such as brain-derived neurotrophic factors (BDNF), which support neuronal survival and function. This production of BDNF may provide further protection against PD. Some occupational activities may also contribute to a reduction in neuroinflammation, a key factor in PD pathogenesis. This effect could be mediated through the direct anti-inflammatory effects of physical activity or through indirect mechanisms involving occupational exposures that modulate inflammatory pathways. Furthermore, occupations promoting physical activity may enhance autophagy, a cellular process crucial for removing damaged proteins and organelles. Efficient autophagy is particularly important in preventing the accumulation of α-synuclein, a hallmark of PD pathology.

**Table 14:** Overview and Mechanisms of Occupations Potentially Reducing the Risk of PD.

| Factor | Key Observations | Mechanisms | References |
| --- | --- | --- | --- |
| **Physical Activity in Occupations** | Jobs involving physical exertion (e.g., engineering, production, construction, and metalworking) reduce PD risk. | Physical activity promotes neuroprotective benefits similar to structured exercise, improving mitochondrial function, reducing oxidative stress, and enhancing neuroplasticity. | [254] |
| **Artistic Occupations** | Engagement in artistic professions (e.g., visual arts, music, writing) correlates with lower PD risk. | Artistic occupations may preserve dopaminergic activity, as creativity requires dopaminergic function, offering neuroprotective effects through sustained brain function. | [255] |
| **Professional/Technical Occupations** | Occupations requiring cognitive engagement and technical skills (e.g., engineering, technical professions) show an inverse relationship with PD risk. | High cognitive engagement may promote neuroplasticity, maintaining neural networks and brain health, which protects against neurodegenerative diseases. | [256] |
| **Farming Activities** | Certain farming activities (e.g., gardening, landscaping, horse-related tasks) have been linked to reduced PD risk. | Physical activity in farming, along with cognitive engagement, may support neuronal resilience, reduce neuroinflammation, and promote neurotrophic factor production like BDNF, which supports neuronal survival. | [257] |

## 4.5 Therapeutic Implications

The state of the gut microbiota significantly influences PD progression by maintaining chronic inflammation and facilitating the spread of misfolded proteins from the gut to the brain. Emerging therapeutic strategies targeting gut dysbiosis aim to restore microbiota balance and potentially slow PD progression. These strategies include probiotics and dietary interventions, which have shown promise in preclinical studies [147, 149].

## 5- Challenges and Solutions in Identifying Risk and Protective Factors for PD

Identifying risk and protective factors for PD is a complex task due to the multifactorial nature of the disease and the intricate interplay between genetic, environmental, and behavioral factors. PD is influenced by a combination of genetic predispositions, environmental exposures, and lifestyle behaviors [257]. Genome-wide association studies (GWAS) have identified over 200 genes associated with PD [19]. However, the genetic landscape is further complicated by rare variants, which are difficult to replicate across studies due to their low frequency [258]. Additionally, the interactions between genetic factors and environmental exposures, such as pesticides and industrial chemicals, remain poorly understood, making it challenging to identify common molecular pathways or therapeutic targets [259]. PD is a heterogeneous disorder with multiple subtypes, each potentially governed by distinct molecular mechanisms [259].This heterogeneity complicates the identification of universal risk or protective factors, as findings relevant to one subtype may not apply to others. Variability in disease progression, shaped by genetic diversity and environmental influences, underscores the need for personalized research approaches and the consideration of population-specific genetic architectures [26].

Several methodological challenges hinder progress in identifying risk and protective factors for PD. Cross-sectional studies, commonly used in PD research, fail to capture the temporal relationships between exposures and disease onset. While longitudinal studies are more informative, they are resource-intensive and less frequently conducted [260]. Confounding factors, such as lifestyle differences, regional variations in environmental exposures, and genetic predispositions, can obscure true associations common[261]. Many studies also face issues like recall bias in case-control designs and difficulties in establishing causality in observational research. Furthermore, inconsistencies across studies—stemming from variations in study design, sample size, and methodologies—impede the development of clear preventive guidelines. For example, while some studies have identified smoking and coffee consumption as protective factors, others have not observed significant associations [261]. Recent advancements in research methodologies offer promising avenues for understanding PD etiology. Techniques such as polygenic risk scores, epigenetic analyses, and Mendelian randomization (MR) provide new opportunities to explore genetic and environmental interactions [259]. However, these approaches require validation and standardization to ensure their broad applicability. The lack of comprehensive data integrating genetic, environmental, and lifestyle factors further limits a holistic understanding of PD risk [261].

To address these challenges, standardizing research methodologies is essential. Implementing uniform protocols for data collection, clinical assessments, and environmental exposure reporting can reduce inconsistencies across studies [262]. Multi-omics approaches, which integrate genomics, proteomics, and metabolomics with artificial intelligence, can uncover patterns that traditional methods might miss, aiding in the identification of biomarkers and elucidating PD's molecular underpinnings. Advanced trial designs, including adaptive trials and studies targeting pre-symptomatic stages, can improve understanding of risk factors and protective measures before significant neuronal damage occurs. Incorporating digital tools and telehealth into research can provide continuous monitoring and overcome limitations associated with traditional clinical endpoints [260].

Large-scale collaborative initiatives are critical to overcoming data fragmentation and leveraging diverse expertise. Establishing centers of excellence, such as those focusing on complex disease genetics (e.g., University of Helsinki), and fostering international collaborations can accelerate data sharing and discovery [262].. Emerging technologies, including next-generation sequencing (NGS), medical digital twins, and blockchain-based data management systems, offer promising pathways to enhance data integration, ensure privacy, and enable personalized research strategies. By addressing these challenges and embracing innovative solutions, the field can progress towards a more comprehensive understanding of PD's etiology, ultimately enabling more effective prevention and therapeutic interventions. Table 15 provides a summary of the challenges and proposed solutions in identifying risk and protective factors for PD.

**Table 15:** Challenges and Proposed Solutions in Identifying Risk and Protective Factors for PD.

| Challenge | Details | Proposed Solutions | References |
|---|---|---|---|
| **Complex Interplay of Genetic and Environmental Factors** | Genetic predispositions interact with environmental exposures, complicating risk factor identification. Over 200 genes are implicated, with rare variants posing replication challenges. | -Integrate multi-omics data (genomics, proteomics, etc.) using AI to uncover patterns. - Identify specific biomarkers through advanced techniques. | [26, 257-259] |
| **Heterogeneity of PD** | PD encompasses multiple subtypes with distinct molecular mechanisms. Variability in progression complicates universal findings. | Employ personalized approaches and population-specific studies. - Account for subtype-specific variability in genetic and environmental influences. | [26, 259] |
| **Methodological Challenges** | Cross-sectional designs fail to capture temporal relationships. Confounding factors like lifestyle and recall bias obscure results. | Conduct longitudinal studies to establish causality. - Standardize protocols for clinical assessments and exposure reporting. Integrate digital tools and telehealth for continuous monitoring. | [260, 261] |
| **Inconsistencies Across Studies** | Variations in study designs, sample sizes, and methodologies lead to contradictory findings. Examples include mixed results for smoking and coffee as protective factors. | Standardize methodologies across studies. - Use collaborative international frameworks for uniformity in research approaches and data collection. | [261] |
| **Evolving Research Methodologies** | Advanced techniques like Mendelian Randomization and polygenic risk scores require validation and standardization. | Validate emerging methods through large-scale studies. Develop global standards to ensure applicability across populations. | [259] |
| **Lack of Comprehensive Data** | Current studies focus on isolated factors, limiting holistic understanding. | Create large-scale collaborative initiatives and centers of excellence for data integration. - Leverage blockchain and digital twins for secure, holistic, and personalized research. | [261, 262] |
| **Technological Advancements** | New tools like next-generation sequencing (NGS) and digital twins offer potential but need broader adoption. | Adopt NGS for genetic analysis and medical digital twins for personalized simulation. [262] Use blockchain for secure data management and sharing. | |

## 6- Raising Awareness in PD

Raising awareness in PD is crucial for understanding its mechanisms, risk factors, and the factors influencing its onset and progression. For healthcare providers, staying informed about PD helps in better managing risk factors like genetics, age, and environmental influences. Highlighting preventive measures such as regular physical activity, balanced nutrition, and early detection is essential to reduce risks and improve patient outcomes. Raising awareness among both professionals and patients can lead to more effective management strategies, ultimately enhancing quality of life and minimizing the impact of the disease. Table 16 outlines recommendations for the prevention and management of Parkinson's Disease, addressing healthy individuals, individuals diagnosed with PD, and healthcare systems.

**Demographic and Genetic Factors**
- **Age**: Aging is a major risk factor for PD, especially among those over 60. Regular screenings (e.g., cognitive assessments, dopamine transporter scans) for neurodegenerative markers are crucial for early detection. Healthy individuals over 60 should consider these preventive measures. PD patients should have regular check-ups to monitor disease progression. Health systems should prioritize neurodegenerative screenings in older adults for early intervention.

- **Sex**: Men are at higher risk for PD. Public health campaigns should target men, focusing on both motor and non-motor symptoms. Healthy men should engage in regular screenings, while those diagnosed with PD need tailored care for gender-specific symptoms.
- **Race and Ethnicity**: Certain racial and ethnic groups have higher PD risk. Targeted health programs and screenings can improve early diagnosis and access to treatments. High-risk populations should participate in community-based screenings, and health systems should implement targeted interventions for these groups.
- **Family History and Genetics**: A family history of PD increases risk. Genetic counseling and predictive testing (e.g., LRRK2 mutations) can help monitor and personalize preventive care. Genetic predisposed individuals might benefit from targeted therapies like LRRK2 inhibitors. Health systems should provide genetic counseling and predictive testing for at-risk families.

**Cognitive and Educational Factors**

Engaging in intellectually stimulating activities (e.g., reading, learning new skills) can reduce the risk of PD progression and support cognitive health. PD patients should incorporate cognitive therapies into their daily routines. Healthcare providers should encourage lifelong cognitive engagement and brain-health-focused educational programs.

**Environmental and Occupational Risks**
- **Toxins**: Exposure to pesticides, heavy metals, and industrial chemicals increases PD risk. Workers should wear protective gear and seek safer alternatives. Regular health check-ups and screenings are vital for those exposed to toxins. Health systems should mandate safety regulations and offer regular screenings for at-risk workers.
- **Air Pollution**: Living in high-pollution areas increases risk. Urban residents should use air purifiers and masks. Public health initiatives should emphasize the connection between air pollution and PD risk and advocate for cleaner air policies.
- **Socioeconomic Disparities**: Limited healthcare access increases PD risk in underserved populations. Expanding preventive screenings and improving access to care in these communities is essential. Health systems should provide affordable screenings and treatments for underserved populations.

**Lifestyle and Health Habits**

- **Nutrition and Diet**: A diet rich in antioxidants, healthy fats, and anti-inflammatory nutrients (e.g., Mediterranean diet) supports brain health. PD patients should focus on neuroprotective foods (e.g., omega-3s, vitamin E). Health systems should encourage nutritional counseling for individuals at risk and PD patients.
- **Physical Activity**: Regular exercise, especially aerobic activities and balance-enhancing exercises, can reduce PD risk. PD patients should follow personalized exercise plans to manage symptoms. Health systems should promote physical activity and integrate physical therapy into PD care.
- **Alcohol Consumption**: Excessive alcohol intake may contribute to neurotoxicity. Healthy individuals should limit alcohol intake to protect brain health. PD patients should monitor alcohol consumption to avoid interactions with medications. Health systems should promote healthy drinking habits and provide resources for alcohol reduction.
- **Body Weight and Metabolic Health**: Maintaining a healthy weight through diet and exercise reduces the risk of neurodegeneration. PD patients should monitor BMI and metabolic health to prevent disease progression. Health systems should prioritize metabolic health screening for at-risk individuals and PD patients.

**Metabolic and Systemic Health Conditions**

- **Diabetes and Hypertension**: Managing blood sugar and blood pressure can reduce PD risk. PD patients should closely monitor these conditions, as they may exacerbate neurological decline. Healthcare providers should emphasize managing diabetes and hypertension in PD care.
- **Cholesterol and Triglycerides**: Maintaining balanced lipid levels through diet and medications can reduce inflammation and protect brain health. PD patients should monitor lipid levels regularly. Health systems should incorporate lipid screenings into PD management and preventive care.
- **Metabolic Syndrome**: Addressing metabolic syndrome with weight management and a healthy lifestyle reduces neurodegenerative risks. PD patients should focus on improving diet and weight management to reduce systemic inflammation. Health systems should offer metabolic screenings for at-risk populations.

**Inflammatory and Immune-Related Factors**

- **Neuroinflammation and Chronic Diseases**: Anti-inflammatory treatments (e.g., NSAIDs, curcumin) may help mitigate PD-related inflammation. Managing chronic inflammatory diseases (e.g., arthritis) can support neurological health. Vaccinations can also reduce infection risk. PD patients should consider anti-inflammatory treatments and manage chronic conditions. Health systems should integrate care for chronic inflammatory diseases and promote vaccination programs.

**Neurological and Psychological Factors**

- **Traumatic Brain Injury (TBI)**: High-risk occupations should enforce safety measures (e.g., helmet use, concussion monitoring) and provide annual neurocognitive assessments.
- **Sleep Disorders**: Screening for sleep disturbances (e.g., REM sleep behavior disorder) is important, as these are linked to increased PD risk. Treatment options like melatonin or CPAP therapy can improve sleep quality.
- **Loss of Smell (Anosmia)**: Anosmia is an early symptom of PD. Routine smell tests in neurological assessments can help detect high-risk individuals.
- **Stress, Depression, and Anxiety**: Chronic stress and mental health issues can accelerate neurodegeneration. Psychological support (e.g., CBT) is essential. Early intervention for depression and anxiety can improve quality of life and may slow PD progression.

**Gut-Brain Axis and Microbiome Health**

Maintaining gut health through a balanced diet with prebiotics and probiotics is essential. PD patients should work with healthcare providers to monitor gut health and explore probiotic therapies. Health systems should facilitate collaboration between neurologists and gastroenterologists to explore gut-brain connections in PD care.

**Hormonal Factors and Other Health Conditions**

- **Estrogen and Neuroprotection**: Postmenopausal women may benefit from hormone replacement therapy (HRT) for neuroprotection. Regular hormonal evaluations should guide treatment.
- **Uric Acid Levels**: Managing uric acid levels through diet and hydration may help lower neurodegenerative risks. PD patients should monitor uric acid levels to reduce neurodegeneration. Health systems should offer regular evaluations for HRT and monitor uric acid levels in PD patients.

**Medications and Emerging Treatments**

- **Neuroprotective Drug Strategies**: Emerging treatments like GLP-1 receptor agonists may reduce PD risk. Healthy individuals at risk should discuss these treatments with healthcare providers. PD patients should explore these drugs and monitor long-term medication effects. Health systems should support research into neuroprotective drugs and monitor long-term medication effects in PD care.

**Table 16:** PD Prevention and Management Recommendations.

| Factor | Recommendation for Healthy People | Recommendation for PD Patients | Health System Consideration |
|---|---|---|---|
| **Age** | Regular screenings for neurodegenerative markers, cognitive assessments | Regular check-ups to monitor disease progression and adjust treatment | Prioritize neurodegenerative screenings for older adults |
| **Sex** | Engage in regular screening, focus on motor and non-motor symptoms | Tailored care for gender-specific risks | Targeted health campaigns focusing on men for early diagnosis |
| **Race and Ethnicity** | Participate in community-based health programs and screenings | Targeted interventions for early diagnosis and treatment access | Implement targeted screenings for high-risk racial and ethnic groups |
| **Family History and Genetics** | Consider genetic counseling and predictive testing (e.g., LRRK2) | Explore targeted therapies for genetically predisposed individuals | Offer genetic counseling and predictive testing for families with PD history |
| **Cognitive and Educational Factors** | Engage in mentally stimulating activities (e.g., reading, memory exercises) | Incorporate cognitive therapies and brain-training exercises | Encourage educational programs and cognitive activities for brain health |
| **Exposure to Toxins** | Use protective gear in high-risk occupations and reduce toxin exposure | Regular consultations to monitor toxin exposure and neurological health | Advocate for safer pesticide use, enforce safety regulations in industries |
| **Air Pollution** | Use air purifiers and masks in high-pollution areas | Focus on reducing exposure to air pollutants | Promote clean air policies and raise awareness about air pollution risks |
| **Socioeconomic Disparities** | Seek affordable preventive screenings | Expand access to healthcare for underserved communities | Offer subsidized screenings and care in underserved areas |
| **Nutrition and Diet** | Adopt a diet rich in antioxidants, healthy fats, and anti-inflammatory foods | Follow a specialized diet with omega-3s, antioxidants, and vitamin E | Encourage nutritional counseling for those at risk and PD patients |
| **Physical Activity** | Engage in aerobic and balance-enhancing exercises | Regular physical therapy and exercise regimens to manage motor symptoms | Promote physical activity and integrate therapy into PD management plans |
| **Alcohol Consumption** | Limit excessive alcohol intake | Consult healthcare providers to monitor alcohol use and interactions | Educate on the risks of excessive alcohol intake and offer reduction resources |
| **Body Weight and Metabolic Health** | Maintain a healthy weight through diet and exercise | Monitor BMI and metabolic health to manage obesity-related risks | Regular metabolic health and BMI screenings as part of preventive care |
| **Diabetes and Hypertension** | Maintain blood sugar levels and blood pressure | Manage blood pressure and glucose levels to support neurological function | Emphasize managing diabetes and hypertension in PD prevention and care |
| **Cholesterol and Triglycerides** | Manage lipid levels with diet and medications | Ensure cholesterol and triglyceride levels are in check to support brain health | Regular lipid screenings as part of PD management |
| **Metabolic Syndrome** | Address with weight management and healthier lifestyle choices | Manage weight and diet to reduce inflammation and support brain health | Provide metabolic screenings and counseling for those at risk |
| **Neuroinflammation and Chronic Diseases** | **Anti-inflammatory treatments:** Consider incorporating anti-inflammatory foods (e.g., curcumin, flavonoids) into the diet to reduce inflammation. | **Anti-inflammatory treatments:** PD patients could benefit from anti-inflammatory medications (e.g., NSAIDs) or dietary supplements to help manage PD-related inflammation. | Promote research into anti-inflammatory treatments and provide guidance on inflammation management for PD patients. |
| **Chronic Inflammatory Diseases & Autoimmune Conditions** | Manage chronic inflammatory diseases like arthritis early to prevent complications that may contribute to neurodegeneration. | Manage autoimmune conditions with appropriate treatments to reduce inflammation and minimize their impact on neurological health. | Provide integrated care between neurologists and specialists in autoimmune diseases to ensure better management of PD-related inflammation. |
| **Preventive Vaccinations** | Get regular vaccinations (e.g., flu shots) to prevent infections that may increase the risk of neurodegeneration. | Preventive vaccinations and antiviral therapies may reduce the risk of infections that can exacerbate PD symptoms. | Ensure access to vaccines and antiviral treatments for PD patients to reduce the likelihood of infection-induced complications. |
| **Traumatic Brain Injury (TBI)** | Follow safety measures in high-risk occupations (e.g., wearing helmets in contact sports or construction). | Monitor for any history of traumatic brain injury, as it may accelerate neurodegeneration in PD patients. | Implement policies for mandatory concussion screening and neurocognitive assessments for individuals in high-risk occupations. |
| **Sleep Disorders** | Prioritize good sleep hygiene and seek medical advice for sleep disturbances. | Screen for sleep disorders like REM sleep behavior disorder and use treatments like melatonin or CPAP therapy to improve sleep quality. | Incorporate sleep disorder screenings into routine neurological assessments for both healthy individuals at risk and PD patients. |
| **Loss of Smell (Anosmia)** | Stay aware of any early signs of anosmia and seek medical consultation if symptoms develop. | Use routine smell tests during neurological assessments to detect high-risk individuals and diagnose early PD. | Offer routine smell tests as part of early PD detection, especially for individuals at risk of developing PD. |
| **Stress, Depression, and Anxiety** | Manage chronic stress through lifestyle changes, including exercise and mindfulness practices. | Engage in psychological support like cognitive behavioral therapy (CBT) and consider medications (e.g., SSRIs) to manage depression and anxiety. | Integrate mental health support services into PD care and emphasize early interventions for stress and depression in both healthy individuals and PD patients. |
| **Gut-Brain Axis and Microbiome Health** | Maintain a healthy gut by consuming a diet rich in prebiotics, probiotics, and fiber to support brain health. | Work with both neurologists and gastroenterologists to identify gut dysbiosis early and incorporate gut health | Promote interdisciplinary collaboration between neurologists and gastroenterologists to monitor and |

|  | | therapies such as probiotics and dietary changes. | manage the gut-brain connection in PD patients. |
|---|---|---|---|
| **Estrogen and Neuroprotection** | Postmenopausal women should consult with their healthcare providers about the potential benefits of hormone replacement therapy (HRT). | Postmenopausal women with PD may benefit from HRT for neuroprotection and should discuss this option with their healthcare providers. | Offer regular hormonal evaluations and consider HRT for postmenopausal women at risk of or diagnosed with PD to support brain health. |
| **Uric Acid Levels** | Maintain healthy uric acid levels through dietary modifications (e.g., reducing purine-rich foods). | Clinicians should monitor uric acid levels in PD patients and consider medication if levels are elevated to reduce neurodegenerative risks. | Regularly monitor uric acid levels in at-risk individuals and PD patients as part of preventive care. |
| **Neuroprotective Drug Strategies** | Consider discussing emerging treatments, such as GLP-1 receptor agonists, with a healthcare provider if at high risk for PD. | For PD patients, emerging drugs like GLP-1 receptor agonists or β2-adrenoceptor antagonists could be considered as part of the treatment plan. | Encourage research into new neuroprotective drug therapies, and integrate emerging treatments into clinical practice for PD prevention and management. |
| **Long-Term Drug Use** | Be aware of the potential neurological effects of long-term drug use and discuss any concerns with healthcare providers. | Monitor the neurological effects of long-term medications and consider safer alternatives if necessary. | provide regular assessments of the long-term effects of medications and offer alternative therapies when available. |

## 7- Conclusion

This paper has provided a comprehensive overview of key risk and prevalent factors that highlight the considerable heterogeneity and complexity underlying the neuropathogenesis of PD. It is likely that multiple factors, with varying degrees of risk, interact within genetically predisposed individuals during critical stages of neurodegeneration, ultimately leading to the clinical manifestation of PD. For each risk and prevalent factor, we discussed the molecular mechanisms involved, which may help to identify the interactive effects of multiple risk factors simultaneously. Additionally, we explored the evolving understanding of risk and prevalence factors in PD, acknowledging that as a complex neurological disorder, new risk factors—particularly those related to genetics—are likely to be uncovered in the future. Furthermore, we highlighted several protective factors that may mitigate the development of PD or slow its progression, such as physical activity. These factors warrant increased attention from individuals and health systems alike to raise awareness and improve public health initiatives within our communities.

**Competing interests**

There are no competing interests declared by the author.

[92] D. Chang et al., "A meta-analysis of genome-wide association studies identifies 17 new Parkinson's disease risk loci," (in eng), *Nat Genet,* vol. 49, no. 10, pp. 1511-1516, Oct 2017, doi: 10.1038/ng.3955.
[93] H. J. Bruce et al., "American football play and Parkinson disease among men," *JAMA Network Open,* vol. 6, no. 8, pp. e2328644-e2328644, 2023.
[94] F. Yang, A. L. V. Johansson, N. L. Pedersen, F. Fang, M. Gatz, and K. Wirdefeldt, "Socioeconomic status in relation to Parkinson's disease risk and mortality: A population-based prospective study," (in eng), *Medicine (Baltimore),* vol. 95, no. 30, p. e4337, Jul 2016, doi: 10.1097/md.0000000000004337.
[95] L. M. Lix, D. E. Hobson, M. Azimaee, W. D. Leslie, C. Burchill, and S. Hobson, "Socioeconomic variations in the prevalence and incidence of Parkinson's disease: a population-based analysis," (in eng), *J Epidemiol Community Health,* vol. 64, no. 4, pp. 335-40, Apr 2010, doi: 10.1136/jech.2008.084954.
[96] A. Ascherio and M. A. Schwarzschild, "Lifestyle and Parkinson's disease progression," (in eng), *Mov Disord,* vol. 34, no. 1, pp. 7-8, Jan 2019, doi: 10.1002/mds.27566.
[97] E. Roos et al., "Body mass index, sitting time, and risk of Parkinson disease," *Neurology,* vol. 90, no. 16, pp. e1413-e1417, 2018.
[98] G. Hu, P. Jousilahti, A. Nissinen, R. Antikainen, M. Kivipelto, and J. Tuomilehto, "Body mass index and the risk of Parkinson disease," *Neurology,* vol. 67, no. 11, pp. 1955-1959, 2006.
[99] K.-Y. Park, G. E. Nam, K. Han, H.-K. Park, and H.-S. Hwang, "Waist circumference and risk of Parkinson's disease," *npj Parkinson's Disease,* vol. 8, no. 1, p. 89, 2022.
[100] G. Hu, P. Jousilahti, A. Nissinen, R. Antikainen, M. Kivipelto, and J. Tuomilehto, "Body mass index and the risk of Parkinson disease," (in eng), *Neurology,* vol. 67, no. 11, pp. 1955-9, Dec 12 2006, doi: 10.1212/01.wnl.0000247052.18422.e5.
[101] K. Cumming, A. D. Macleod, P. K. Myint, and C. E. Counsell, "Early weight loss in parkinsonism predicts poor outcomes: evidence from an incident cohort study," *Neurology,* vol. 89, no. 22, pp. 2254-2261, 2017.
[102] S. Peters et al., "Alcohol consumption and risk of Parkinson's disease: data from a large prospective European cohort," *Movement disorders,* vol. 35, no. 7, pp. 1258-1263, 2020.
[103] C. Shao, X. Wang, P. Wang, H. Tang, J. He, and N. Wu, "Parkinson's Disease Risk and Alcohol Intake: A Systematic Review and Dose-Response Meta-Analysis of Prospective Studies," *Frontiers in Nutrition,* vol. 8, p. 709846, 2021.
[104] F. J. Jiménez-Jiménez, H. Alonso-Navarro, E. García-Martín, and J. A. Agúndez, "Alcohol consumption and risk for Parkinson's disease: a systematic review and meta-analysis," *Journal of neurology,* vol. 266, no. 8, pp. 1821-1834, 2019.
[105] R. Kelsey, "Lifestyle factors and progression of PD," *Nature Reviews Neurology,* vol. 15, no. 3, pp. 126-126, 2019.
[106] R. Daviet et al., "Associations between alcohol consumption and gray and white matter volumes in the UK Biobank," *Nature Communications,* vol. 13, no. 1, p. 1175, 2022.
[107] C. Domenighetti et al., "Dairy intake and Parkinson's disease: a mendelian randomization study," *Movement Disorders,* vol. 37, no. 4, pp. 857-864, 2022.
[108] K. C. Hughes et al., "Intake of dairy foods and risk of Parkinson disease," *Neurology,* vol. 89, no. 1, pp. 46-52, 2017.
[109] H. Chen et al., "Consumption of dairy products and risk of Parkinson's disease," *American journal of epidemiology,* vol. 165, no. 9, pp. 998-1006, 2007.